\documentclass[%
reprint,
amsmath,amssymb,
aps,
prd,
]{revtex4-2}

\usepackage{ORCIDinREVTEX}
\usepackage{graphicx}
\usepackage{dcolumn}
\usepackage[caption=false]{subfig}
\usepackage{lmodern} 
\usepackage{courier} 
\usepackage{natbib}
\usepackage{bm}
\usepackage{hyperref}
\hypersetup{colorlinks=true, citecolor=cyan, urlcolor = blue, linkcolor = blue}

\usepackage{aas_macros}
\usepackage{color}
\newcolumntype{S}{>{\centering\arraybackslash} m{.6\linewidth} }

\begin{document}

\title{Fast wavelet basis search for generic gravitational wave bursts in Pulsar Timing Array data}
\author{Jacob A.~Taylor}
\orcid{0000-0001-9118-5589}
\author{Rand Burnette} 
\orcid{0009-0008-3649-0618}
\author{Bence B\'{e}csy}
\orcid{0000-0003-0909-5563}
\affiliation{%
Department of Physics, Oregon State University, Corvallis, OR 97331, USA
}
\author{Neil J.~Cornish}
\orcid{0000-0002-7435-0869}
\affiliation{%
 eXtreme Gravity Institute, Department of Physics, Montana State University, Bozeman, MT 59717, USA
}%
%


\date{\today}

\begin{abstract}
As we move into an era of more sensitive pulsar timing array data sets, we may be able to resolve individual gravitational wave sources from the stochastic gravitational wave background. While some of these sources, like orbiting massive black hole binaries, have well-defined waveform models, there could also be signals present with unknown morphology. This motivates the search for generic gravitational-wave bursts in a signal-agnostic way. However, these searches are computationally prohibitive due to the expansive parameter space. In this paper we present \texttt{QuickBurst}, an algorithm with a re-defined likelihood that lets us expedite Markov chain Monte Carlo sampling for a subset of the signal parameters by avoiding repeated calculations of costly inner products. This results in an overall speedup factor of $\sim$200 on realistic simulated datasets, which is sufficient to make generic gravitational-wave burst searches feasible on current and growing pulsar timing array datasets.
\end{abstract}

\maketitle


\section{\label{sec:intro} Introduction} 

The North American Nanohertz Observatory for Gravitational Waves (NANOGrav), the Parkes Pulsar Timing Array (PPTA), the European Pulsar Timing Array (EPTA), and the Chinese Pulsar Timing Array (CPTA) have found evidence for a stochastic gravitational-wave (GW) background (GWB) with varying level of significance \cite{nanograv_15_gwb, epta_dr2_gwb, ppta_dr3_gwb, cpta_dr1_gwb}. This is exciting for both multimessenger astrophysics and future prospects for detecting individual binary systems \cite{nanograv_12p5_cw, nanograv_15_cw, BayesHopperBurst}. In addition to the persistent stochastic signal of the GWB, there are many other possible sources of short-duration GW bursts. Examples are transient deterministic signals whose persistence is less than the dataset time span, and these signals have the potential to be observed via pulsar timing arrays (PTAs) \cite{12p5yr_NG_BWM,  nanograv_15yr_newphysics}. In order to agnostically search for these potential sources, we can use a generic GW burst model to find transient GWs with a variety of waveforms. 

Previous searches have been done for gravitational wave burst signals in LIGO data (see e.g.~\cite{LIGO_o3_short_burst, LIGO_o3_long_burst}) using various techniques \cite{BayesWaveI, BayesWaveII, BayesWaveIII, cWB, oLIB}. Various techniques have also been proposed in other GW frequency bands, including LISA \cite{Robson_Cornish_LISA_burst} and PTA data \cite{Zhu_et_al_burst, BayesHopperBurst, Heling_pwl}. One prominent approach is to model the waveform with a collection of wavelets, where the number of these wavelets used is also a free model parameter, allowing for flexible signal modeling. This previously proposed wavelet-based technique would take months to perform on PTA datasets in comparison to $\sim$ week time-frame it takes to do a single analysis run for GWB searches on cutting edge hardware. This is because of the large parameter space needed to fully model a GW burst in PTA data. To feasibly do a generic GW burst search on current and future data sets, we need to improve the efficiency of our generic GW burst search algorithms. 

In this paper, we present \texttt{QuickBurst}\footnote{Code publicly available at: \url{https://github.com/JacobATaylor/QuickBurst}}, a Bayesian generic GW burst search algorithm that consists of a trans-dimensional reversible-jump Markov chain Monte Carlo (RJMCMC) sampler \cite{rjmcmc, rjmcmc_new} and a factorized likelihood template that utilizes the properties of a deterministic signal model for improved computational efficiency. Recent work has shown that for some deterministic signals, a faster search algorithm can be achieved by pre-calculating and storing components of the likelihood calculation that do not change \cite{QuickCW}. The work presented here is built upon \texttt{BayesHopperBurst} and \texttt{QuickCW}, the algorithms presented in \cite{BayesHopperBurst} and \cite{QuickCW} respectively. \texttt{BayesHopperBurst} uses an RJMCMC sampler to search for generic GW burst signals, but does not utilize more efficient sampling methods, of which their usefulness has been demonstrated in \texttt{QuickCW}.

In \S\ref{sec:Quick_like}, we derive the Bayesian likelihood for a generic GW burst signal modeled as a superposition of Morlet-Gabor wavelets. We also describe the techniques employed in our Markov chain Monte Carlo (MCMC) sampling algorithm that take advantage of our likelihood formulation. Then, we present the expected computational efficiency improvements that these techniques produce, as well as our verification tests to ensure the operational quality of the algorithm. In \S\ref{sec:Analysis_of_Datasets}, we demonstrate the capabilities of \texttt{QuickBurst} by performing searches on simulated datasets that resemble the NANOGrav 15-year dataset \cite{NG15_data}. Furthermore, we test the effects of a common uncorrelated red noise (CURN) process on \texttt{QuickBurst}'s ability to recover a GW burst signal. Finally, in \S\ref{sec:Conclusion} we discuss future applications involving \texttt{QuickBurst} to search for gravitational wave burst signals in PTA data.

\section{\label{sec:Quick_like} QuickBurst Methods} 
\subsection{Models} \label{Models}
We can model the timing residuals for the PTA as a vector

\begin{equation} \label{eq:PTA_residuals}
\delta \textbf{t} = \begin{bmatrix}
    \delta \textbf{t}_{1}\\
    \delta \textbf{t}_{2}\\
    \vdots\\
    \delta \textbf{t}_{\mathrm{N_{psr}}} \end{bmatrix},
\end{equation}
where $\mathrm{N_{psr}}$ is the number of pulsars in the PTA. The timing residuals in the $i^{th}$ pulsar is then modeled as
\begin{equation}\label{eq:residuals}
    \delta \boldsymbol{t}_{i} =  M_{i}\delta \boldsymbol{\xi}_{i} + \boldsymbol{n}_{i} + \boldsymbol{g}_{i} + \boldsymbol{\omega}_{i}(\boldsymbol{\theta}_{s}) + \boldsymbol{V}_{i}(\boldsymbol{\theta}_{n}).
\end{equation}

In this formalism, $M_{i}$ is our design matrix with dimensions ($\mathrm{N_{TOA}} \times \mathrm{N_{par}}$), where $\mathrm{N_{TOA}}$ represents the number of times-of-arrival (TOAs) in the $i^{th}$ pulsar and $\mathrm{N_{par}}$ represents the number of timing model parameters in this pulsar, $\delta \boldsymbol{\xi}_{i}$ is a vector of shape $\mathrm{N_{par}}$ that accounts for a small error in this pulsar's timing model, $\boldsymbol{n}_{i}$ represents intrinsic noise in this pulsar, $\boldsymbol{g}_{i}$ represents the contribution from spatially uncorrelated common to all pulsars, $\boldsymbol{\omega}_{i}$ is the contribution from a generic gravitational wave burst signal, and $\boldsymbol{V}_{i}$ represents the contribution to this pulsar's timing residuals of any incoherent noise transients that are unique to this pulsar. $\boldsymbol{\theta}_{s}$ and $\boldsymbol{\theta}_{n}$ are the parameters that describe the GW burst signal and transient noise respectively. $\delta \boldsymbol{t}_{i}$, $\boldsymbol{n}_{i}$, $\boldsymbol{g}_{i}$, $\boldsymbol{\omega}_{i}$, and $\boldsymbol{V}_{i}$ are all vectors with dimensions of $\mathrm{N_{TOA}}$. 

Morlet-Gabor wavelets (sine-Gaussian wavelets) are a robust basis for generically modeling transient signals. First, the transient noise $\boldsymbol{V}_{i}$ may be written as a linear combination of these wavelets:
\begin{equation}\label{eq:transients}
    \boldsymbol{V}_{i}(\boldsymbol{\theta}_{n}) = \sum_{j=1}^{N_{i}} \boldsymbol{\Psi}(\boldsymbol{t}_{i}; \boldsymbol{\lambda}_{ij}),
\end{equation}
where $N_{i}$ is the number of noise transient wavelets in the $i^{th}$ pulsar, $\boldsymbol{t}_{i}$ is the vector of timing observations for the $i^{th}$ pulsar, and $\boldsymbol{\lambda}_{ij}$ are the parameters for the $j^{th}$ wavelet for the $i^{th}$ pulsar. Each wavelet can be expressed as a vector 

\begin{equation}
\label{eq:Noise_transient_vec}
    \boldsymbol{\Psi}(t_{i}; \boldsymbol{\lambda}_{ij}) =  \begin{bmatrix} 
    \Psi(t_{i1}; \boldsymbol{\lambda}_{ij})\\
    \Psi(t_{i2}; \boldsymbol{\lambda}_{ij})\\
    \vdots\\
    \Psi(t_{i \mathrm{N_{TOA}}}; \boldsymbol{\lambda}_{ij}) \end{bmatrix}, 
\end{equation}
where the elements in this vector are described by
\begin{equation}
\label{eq:Morlet_gabor}
    \Psi(t_{i}; \boldsymbol{\lambda}_{ij}) = Ae^{(t_{in}-t_{0})^{2}/{\tau^{2}}} \cos(2\pi f_{0}(t_{in}-t_{0}) + \phi_{0}).
\end{equation}
In this wavelet formulation, $A$ is an overall amplitude, $t_{in}$  is the $n^{th}$ observation time in the $i^{th}$ pulsar, $t_{0}$ is the central time for the wavelet, $\tau$ is the characteristic duration, $f_{0}$ is the central frequency, and $\phi_{0}$ is the initial phase. 

For the signal model, the contributions from a GW burst may also be written as a sum of these Morlet-Gabor wavelets further modulated by the antenna response of the PTA based on GW polarization and source sky location. Because GWs perturb spacetime based on both GW polarization and propagation direction, we sum together our sine-Gaussian wavelets for the signal, and project the signals onto the lines of sight of the pulsars in our array, which are dependent on the corresponding antenna responses and ensures coherence between the pulsars in our array. Quantitatively, we can write the photon time integral of the metric perturbation as a linear combination of wavelets

\begin{align}
\boldsymbol{H}_{+} = \sum_{j=1}^{N} \boldsymbol{\Psi}(t;t_{0,j}, f_{0,j}, \tau_{j}, A_{+,j}, \phi_{0, +, j}), \label{eq:H_plus_template} \\
\boldsymbol{H}_{\times} = \sum_{j=1}^{N} \boldsymbol{\Psi}(t;t_{0,j}, f_{0,j}, \tau_{j}, A_{\times,j}, \phi_{0, \times, j}) \label{eq:H_cross_template}, 
\end{align}
where $\boldsymbol{H}_{+}$ and $\boldsymbol{H}_{\times}$ are the plus and cross GW polarizations respectively, and have independent amplitudes $A_{+}$ and $A_{\times}$, with independent phases $\phi_{0,+}$ and $\phi_{0,\times}$. This leaves $t_{0}, f_{0}$, and  $\tau$ as common parameters between both polarizations. Furthermore, we model the phases independently, as not all GWs have the same 90$^{\circ}$ phase shift relationship between the plus and cross polarizations.


Next, we perform a rotation around the propagation direction of the GW, which is given by
\begin{equation}\label{eq:H_cross_rot}
    \bar{\boldsymbol{H}}_{\times} = \boldsymbol{H}_{\times}\cos(2\psi_{\rm gw}) - \boldsymbol{H}_{\times}\sin(2\psi_{\rm gw}),
\end{equation}
\begin{equation}\label{eq:H_plus_rot}
    \bar{\boldsymbol{H}}_{+} = \boldsymbol{H}_{+}\sin(2\psi_{\rm gw}) + \boldsymbol{H}_{+}\cos(2\psi_{\rm gw}),
\end{equation}
where $\psi_{\rm gw}$ is the polarization angle of the GW. Finally, the contribution of the signal to the timing residual can be expressed as

\begin{equation}
\begin{split}\label{eq:Signal_res}
    \boldsymbol{\omega}_{i}(\theta_{s})  =  &-F_{+}(\Omega_{i}, \Omega_{\rm gw})\bar{\boldsymbol{H}}_{+}(\boldsymbol{t}_{i};\boldsymbol{\lambda}')\\
    &-F_{\times}(\Omega_{i}, \Omega_{\rm gw})\bar{\boldsymbol{H}}_{\times}(\boldsymbol{t}_{i};\boldsymbol{\lambda}'),  
\end{split}
\end{equation}
\\
where $F_{+}(\Omega_{k}, \Omega_{\rm gw})$ and $F_{\times}(\Omega_{k}, \Omega_{\rm gw})$ are the plus and cross polarization mode antenna responses, respectively. For further details on the geometric response, see \cite{Ellis_Antenna}.
Eq.~(\ref{eq:Signal_res}) assumes that the only contribution to the signal is from the Earth term (ET) and not the Pulsar Term (PT). This is a result of our current observing baseline ($\sim20$ years) being significantly shorter than the light travel time between the Earth and any pulsar in the PTA ($\sim 10^{3}$ years). Thus, any GW bursts which appear in only one pulsar will not affect the residuals in any other pulsar (unless two or more pulsars are within a few degrees of each other), nor will it reach the Earth within the time frame of the experiment. As such, this burst will appear as a single noise transient in that particular pulsar. For more details on the geometric time delay between the ET and the PT, see Section 2.1 in Ref. \cite{BayesHopperBurst}.

Then, for $N_{p}$ pulsars, the ET signal-to-noise ratio (SNR) is larger than the PT SNR by a factor of $\sqrt{N_{p}}$. The SNR scaling for the ET also becomes larger when considering that an ET search will allow us to see a common signal in all pulsars in our PTA, while the PT search will only show a signal in one pulsar and can be modeled as a noise transient. For additional details on these arguments, see \cite{BayesHopperBurst, ET_only_arg}. Moreover, it is expected that the number of events that we can possibly observe in the ET only search scales with the number of pulsars $\sim N_{p}^{3/2}$. However, the pulsar term (PT) only search scales as $\sim N_{p}$ due to the extended timing baseline when combining individual pulsar data. As a result, the number of possible observed events in the ET only search exceeds the number of possible observed events in the PT only search by $\sim N_{p}^{1/2}$. Considering current PTAs contain of order 100 pulsars, there's potential to observe $\sim 10\times$ more events in the ET only search.

\vspace{-10pt}
\subsection{Likelihood}
\label{ssec:Likelihood}
We use a factorized likelihood, which allows us to simplify some of our likelihood calculations depending on which parameters in our model we want to vary. We can demonstrate how to generically rewrite a standard likelihood given a specific generic GW burst model template. For any signal, the likelihood can be written as

\begin{equation}
    \begin{aligned}
    \label{eq:LogL_template}
    \log L = \sum_{i = 1}^{N_{\mathrm{psr}}}&\left[-\frac{1}{2} (\delta \boldsymbol{t}_{i} - \boldsymbol{s}_{i}|\delta \boldsymbol{t}_{i} - \boldsymbol{s}_{i}) \right. \\
    & \hspace{4pt}- \left. \frac{1}{2} \log{\det(2\pi C_{i})}\right],
    \end{aligned}
\end{equation}
where $\boldsymbol{s}_{i}$ is our GW burst signal and $C_{i}$ is the noise covariance matrix for the $i^{th}$ pulsar. Because we assume that there are no spatially correlated signals, we can guarantee that the full PTA covariance matrix is block-diagonal. Thus, we can separate the individual pulsar covariance matrices. We define a noise-weighted inner product between any two vectors $\boldsymbol{\mathrm{a}}$ and $\boldsymbol{\mathrm{b}}$ as  

\begin{equation}
\label{eq:dot}
\mathrm{(\boldsymbol{\mathrm{a}}|\boldsymbol{\mathrm{b}})} = \boldsymbol{\mathrm{a}}^{T} \mathrm{C^{-1}} \boldsymbol{\mathrm{b}}, 
\end{equation}
where 
\begin{equation}\label{eq:covariance_matrix}
C^{-1} = N^{-1} - N^{-1} T \Sigma^{-1} T^{T} N^{-1}
\end{equation}
is the inverse of the covariance matrix for a particular pulsar given by the Woodbury identity \cite{Woodbury}, and 

\begin{equation}\label{eq:sigma_inv}
   \Sigma = B^{-1} + T^{T} N^{-1} T.
   \vspace{6pt}
\end{equation}
B is the prior matrix for the hyper-parameters, and T is the design matrix for the timing model, red noise (RN) and jitter (see \cite{Steve_Nanohertz}). The design matrix includes additional models for dispersion measure (DM), specifically modeling DM as an epoch-by-epoch DM offset. This model is refered to as DMX \cite{Steve_Nanohertz}.

For the PTA, we can write the vector of GW burst contributions to the timing residuals in eq.~(\ref{eq:PTA_residuals}) as 

\begin{equation} \label{eq:PTA_signal_model}
\boldsymbol{\mathrm{s}} = \begin{bmatrix}
    \boldsymbol{s}_{1}\\
    \boldsymbol{s}_{2}\\
    \vdots\\
    \boldsymbol{s}_{\mathrm{N_{psr}}} \end{bmatrix},
\end{equation}

where we decompose the signal $\mathrm{\boldsymbol{s}_{i}}$ in the $i^{th}$ pulsar as a sum of filter functions $\boldsymbol{\mathrm{S}}^{\mathrm{ki}} (\theta_{s})$ multiplied by coefficients $\mathrm{\sigma_{ki} (\theta_{p})}$, 

\begin{equation}
\label{eq:signal_template}
    \boldsymbol{s}_{i} = \sum_{k=1,2} \mathrm{\sigma_{ki}} (\theta_{p}) \boldsymbol{\mathrm{S}}^{\mathrm{ki}} (\theta_{s}).
\end{equation}

Direct substitution of eq.~(\ref{eq:signal_template}) into eq.~(\ref{eq:LogL_template}) yields 
\vspace{3pt}
\begin{widetext}
    \begin{equation}
    \label{eq:logL}
    \log L = \sum_{i = 1}^{N_{\mathrm{psr}}}\left[-\frac{1}{2} (\delta \boldsymbol{t}_{i}|\delta \boldsymbol{t}_{i}) - \frac{1}{2} \log{\det(2\pi C_{i})} + \sum_{v = 1}^{N_{F}}\big[\sum_{k=1,2} \sigma_{ki} N_{i}^{k} - \frac{1}{2}\sum_{k=1,2}\sum_{l=1,2} \sigma_{ki} \sigma_{li} M_{i}^{kl}\big]\right],
    \end{equation}
\end{widetext}

where the last two terms are inner products between the $i^{th}$ pulsar residuals and the filter functions for that pulsar, as well as the inner product between the pairs of filter functions respectively. These inner products are defined as
\begin{equation}\label{eq: N_matrix}
   N_{i}^{k} = (\delta \boldsymbol{t}_{i}|\boldsymbol{\mathrm{S}}^{\mathrm{ki}}),
\end{equation}
and 
\begin{equation}\label{eq: M_matrix}
    M_{i}^{kl} = (\boldsymbol{\mathrm{S}}^{\mathrm{ki}}|\boldsymbol{\mathrm{S}}^{\mathrm{li}}),
\end{equation}

where a filter function $\boldsymbol{\mathrm{S}}^{\mathrm{ki}}$ is a vector of size $\mathrm{N_{\rm TOA}}$ (which has the same structure as $\delta \boldsymbol{t}_{i}$). Lastly, $N_{F}$ is the number of wavelets present in the signal model.

Now, we combine eq.~(\ref{eq:H_plus_template}), eq.~(\ref{eq:H_cross_template}), eq.~(\ref{eq:H_cross_rot}), eq.~(\ref{eq:H_plus_rot}), and eq.~(\ref{eq:Signal_res}) to get the filter functions and filter coefficients for the noise transients and the GW burst signal based on eq.~(\ref{eq:signal_template}). In doing this, we get:

\begin{align}
\mathrm{S}^{\mathrm{1i}}_{n} &= e^{(t_{in}-t_{0})^{2}/{\tau^{2}}} \cos(2\pi f_{0}(t_{in}-t_{0})),\\
\mathrm{S}^{\mathrm{2i}}_{n} &= e^{(t_{in}-t_{0})^{2}/{\tau^{2}}} \sin(2\pi f_{0}(t_{in}-t_{0})),
\end{align}

which are the elements of our filter functions $\boldsymbol{\mathrm{S}}^{\mathrm{ki}}(\theta_{s})$ from eq.~(\ref{eq:signal_template}) for the $\emph{i}^{th}$ pulsar and the $n^{th}$ TOA, and the corresponding signal coefficients for a single wavelet are given by

\begin{widetext}
    \begin{equation}
    \label{eq:signal_coeff}
        \begin{aligned}
            \sigma_{1i} &= -\cos(2\psi_{\rm gw})[F_{i,+}A_{+}\cos(\phi_{0,+})+F_{i,\times}A_{\times}\cos(\phi_{0,\times})]+\sin(2\psi_{\rm gw})[F_{i,+}A_{\times}\cos(\phi_{0,\times})-F_{i,\times}A_{+}\cos(\phi_{0,+})],\\
            \sigma_{2i} &= \cos(2\psi_{\rm gw})[F_{i,+}A_{+}\sin(\phi_{0,+})+F_{i,\times}A_{\times}\sin(\phi_{0,\times})]+\sin(2\psi_{\rm gw})[F_{i,\times}A_{+}\sin(\phi_{0,+})-F_{i,+}A_{\times}\sin(\phi_{0,\times})].
        \end{aligned}
    \end{equation}
\end{widetext}

The corresponding noise transient coefficients for a single wavelet are given by

\begin{equation}
    \begin{aligned}
    \sigma_{\textbf{1i}} &= A\cos(\phi_0),\\
    \sigma_{\textbf{2i}} &= -A\sin(\phi_0).
    \end{aligned}
\end{equation}

We can now see why writing this factorized likelihood opens the door for more efficient calculations. For the signal model, we have separated all of the terms that are inherently tied to the pulsar TOA data into the $\mathrm{N_{i}^{k}}$ and $\mathrm{M_{i}^{kl}}$ terms, so the coefficients can be recalculated without doing any of the costly inner products in these matrices. On the other hand, the coefficients, which are cheap to recalculate, can be independently calculated during runtime.

Based on eq.~(\ref{eq:signal_coeff}), we can now define our sets of shape and projection parameters, which are parameters that are either non-separable or separable from pulsar TOAs, respectively.
For noise transients, based on eq.~(\ref{eq:Morlet_gabor}), our shape parameters ($\boldsymbol{\theta}_{s}$) and projection parameters ($\boldsymbol{\theta}_{p}$) are
\begin{equation}
\label{eq:transient_params}
    \begin{aligned}
        \boldsymbol{\theta}_{s} &\rightarrow (\tau , t_{0}, f_{0}),\\
        \boldsymbol{\theta}_{p} &\rightarrow (A, \phi_{0}).
    \end{aligned}
\end{equation}
For our generic GW burst signal, our shape and projection parameters are
\begin{equation}
\label{eq:signal_params}
    \begin{aligned}
        \boldsymbol{\theta}_{s} &\rightarrow (\tau , t_{0}, f_{0}),\\
        \boldsymbol{\theta}_{p} &\rightarrow (\theta_{\rm gw}, \phi_{\rm gw}, \psi, A_{+}, A_{\times},\phi_{0, +}, \phi_{0, \times}).
    \end{aligned}
\end{equation}

We can see from eq.~(\ref{eq:transient_params}), the number of shape and projection parameters for noise transient wavelets in our model go as $3\times N_{n}$ and $2\times N_{n}$, respectively for $N_{n}$ noise transient wavelets. From eq.~(\ref{eq:signal_params}), the number of shape and projection parameters in our model will scale as $3\times N_{s}$ and $3+4\times N_{s}$ respectively for $N_{s}$ GW signal wavelets, since all wavelets share the same sky location and polarization angle. Notably, in previous work using the signal template defined by \cite{QuickCW} for continuous waves, there are always the same number of shape and projection parameters. Our model has more projection than shape parameters, which will mean a larger percent of our model will be calculated with the more efficient methods in our algorithm.

Conveniently, the shape parameters are the same for noise transients and our signal, which necessitates fewer recalculations during sampling. Due to GW information, however, the coefficients $\sigma_{ki}$ depend on the antenna patterns and GW polarization, making them more tedious to recalculate.

\subsection{Sampling} \label{ssec:Sampling}

The likelihood factorization that results from utilizing our signal template, as shown in eq.~(\ref{eq:logL}), allows us to avoid recalculating components of the likelihood depending on what parameters we are varying in any given step. These correspond to the two aforementioned types of parameters, shape parameters and projection parameters, as described in eq.~(\ref{eq:transient_params}) and eq.~(\ref{eq:signal_params}). Sampler steps that only change projection parameters don't require recalculating any components of $\mathrm{N_{i}^{k}}$ and $\mathrm{M_{i}^{kl}}$. We call these projection parameter updates "fast jumps" because the coefficients are several orders of magnitude faster to recalculate than the filter functions. This results in these steps being faster than all other parameter jumps we can do, which vary over various subsets of shape parameters (see Table \ref{tab:timing_table}).

We use similar jump techniques as \texttt{BayesHopperBurst} \cite{BayesHopperBurst} when exploring intrinsic pulsar RN, white noise (WN), or a CURN process. It is worth noting that these do require recalculating more pieces of the likelihood than signal parameter jumps. Based on eq.~(\ref{eq:logL}), our options for parameter types to vary are (in order of decreasing computational expense):
\begin{enumerate}
    \item \textbf{Intrinsic pulsar noise and CURN}: All terms in the likelihood are recalculated as a result of the noise covariance matrix $C$ changing. 
    \item \textbf{Shape parameters}: Parts of the terms which include our signal model need to be recalculated when varying shape parameters. We only recalculate the elements of $M_{i}$ and $N_{i}$ that are changing, while all other elements are stored for the next likelihood calculation.
    \item \textbf{Projection parameters}: Only the filter function coefficients are recalculated. 
\end{enumerate}

We also need to allow the dimensions of our parameter space to change during sampling depending on how many GW signal wavelets or transient wavelets the model prefers. To properly explore these models, we apply trans-dimensional proposals to probe more or less signal or noise transient parameters \cite{rjmcmc, rjmcmc_new} as the sampler evolves. These trans-dimensional proposal steps are labeled as TD in Table \ref{tab:timing_table}. Additionally, we apply Fisher matrix proposals and $\tau$-scan proposals to increase sampling efficiency.

For Fisher matrix proposals, we can numerically construct a matrix from the covariance between our model parameters by computing second derivatives of the likelihood via finite differences. Then, the eigenvectors of the Fisher matrix can be used to change model parameters between MCMC steps based on the covariance between these model parameters. These Fisher matrices are constructed for various sets of model parameters separately from each other (i.e. for intrinsic pulsar noise, signal and noise transient wavelets, CURN, etc.), which allows us to make more informed MCMC jump proposals based on the type of step taken during sampling. Additionally, these fisher matrices can be recalculated during runtime based on the location of the MCMC sampler in parameter space to better inform our Fisher steps during sampling. For more details on how these matrices are computed, see \cite{Fisher_matrix}.

Additionally, we can compute a global proposal for all wavelets to draw from in what we call $\tau$-scan proposals. These proposals are informed by a 3D map over our shape parameters (see eq.~(\ref{eq:transient_params}), eq. ~(\ref{eq:signal_params})) where we find some excess power in our data. $\tau$-scan proposals have been used previously in \texttt{BayesHopperBurst}, and can be utilized in our algorithm to improve sampling efficiency specifically in our shape parameters. More details on $\tau$-scan proposals can be found in \S2 of \cite{BayesHopperBurst}.

\subsection{Verification and Timing \label{ssec:Ver&Timing}}

\begin{figure*}[htbp]
    \centering
    \includegraphics[width = 1.2\columnwidth]{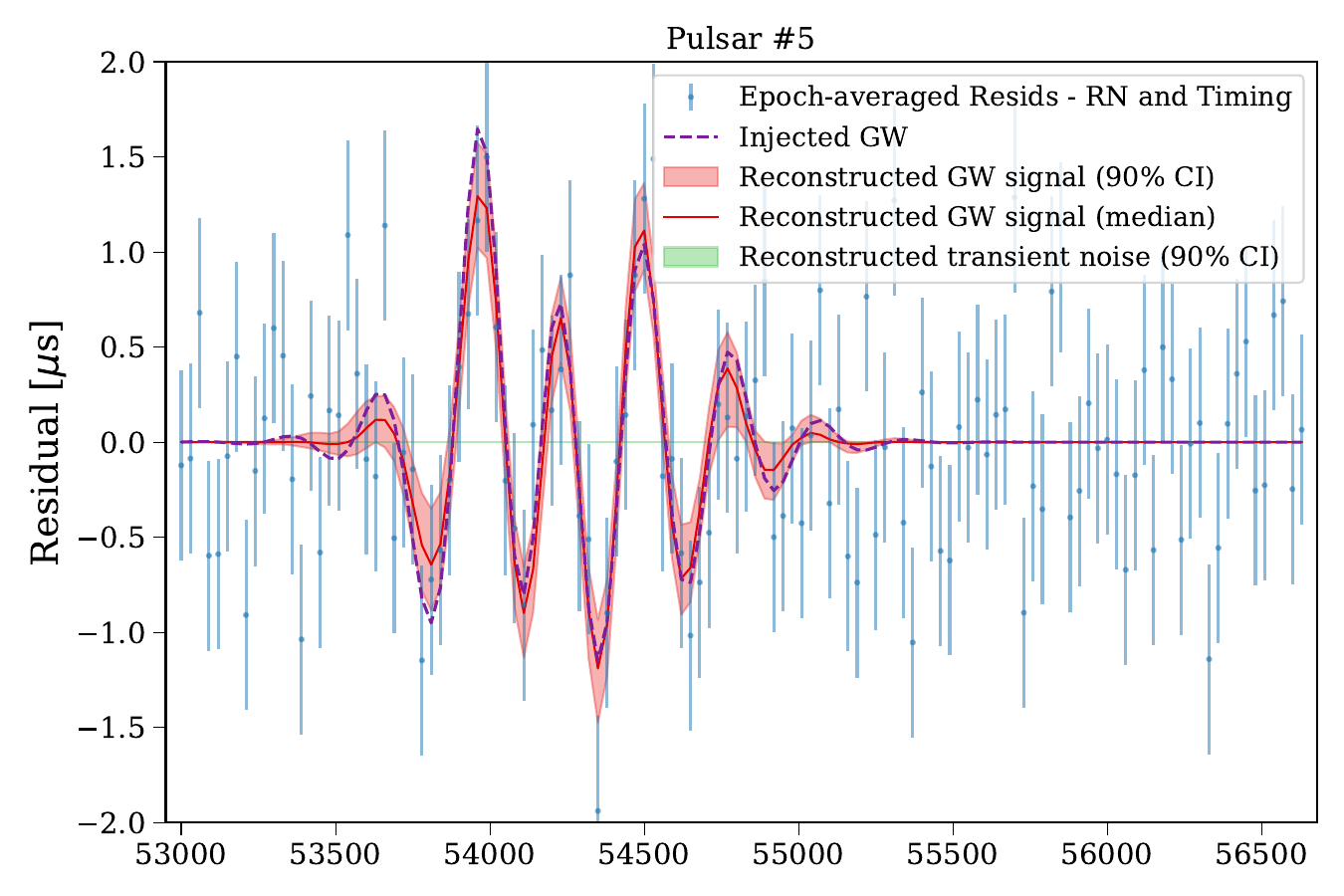}
    \caption{Waveform reconstructions for a simulated dataset containing two sine-Gaussian signal wavelets with an SNR of 17.9. The recovered waveform has an SNR of 16.3, and a match of M = 0.97 with the injected signal (see eq.~(\ref{eq:match_stat})).}
    \label{fig:2wavelet_reconstruction}
\end{figure*}

Before analyzing efficiency improvements, we first present comparisons between the Bayesian likelihood implemented in \texttt{QuickBurst} and \texttt{BayesHopperBurst}. As \texttt{BayesHopperBurst} utilizes the software package \texttt{enterprise} \cite{NG15_methods} to evaluate the Bayesian likelihood, we use these values as a benchmark for our algorithm. The likelihood comparisons are evaluated using a simple data set consisting of two sine-Gaussian wavelets. Our comparisons show that the two signal models yield likelihoods that are equivalent up to $10^{-10}$ precision. When including either intrinsic pulsar red noise or CURN, compounding numerical errors in likelihood evaluations increase differences to $\sim10^{-5}$. However, the spread in likelihood differences is still centered around a difference of zero. Additionally, the signal reconstruction for a single pulsar from this run can be seen in \autoref{fig:2wavelet_reconstruction}.

With consistency between \texttt{enterprise} and \texttt{QuickBurst} established, we can now proceed to demonstrating the improved computational efficiency of \texttt{QuickBurst}. To do so, we simulate datasets as close to the 15 year NANOGrav dataset as possible (see \S \ref{sec:Analysis_of_Datasets} for simulation details) and time our various types of steps. The timing for these steps correspond to runs on the SIM-MID dataset detailed in \autoref{tab:sim_params}, which only vary our GW signal and noise transient parameters. However, in order to time noise jumps, we allow intrinsic pulsar RN to vary. It is worth noting that the timing runs do not include CURN, which is handled in the "Regular Fisher" jump proposals. The average timing for these jumps is around 0.51 s if CURN is included. The results can be seen in Table \ref{tab:timing_table}.

\begin{scriptsize}
\begin{table}[ht!]
    \begin{tabular}{|c|c|c|c|}\hline
    \centering
    \texttt{\textbf{Jump types}}& \texttt{\textbf{QuickBurst}} & \texttt{\textbf{BayesHopperBurst}} \\
    & (s) & (s) \\
    \hline \hline 
    TD Signal  &$ 0.50\ $ & $0.20\ $ \\ \hline 
    TD Noise Transient &$ 0.41\ $ & $0.18\ $ \\ \hline
    Signal $\tau$ Scan & $0.30\ $ & $0.21\ $ \\ \hline
    Noise Transient $\tau$ Scan & $0.24\ $ & $0.18\ $ \\ \hline
    Regular Fisher & $0.23\ $ &$ 0.13\ $ \\ \hline
    Noise & $0.66\ $ &$ 0.08\ $ \\ \hline
    Fast & $1.5 \times 10^{-5}\  $ & \texttt{N/A} \\ \hline \hline
    \centering
    \textbf{\normalsize Total Speed Up} & \multicolumn{2}{| c |}{\rule{0pt}{10pt}\normalsize \textbf{$\boldsymbol{\times}\mathbf{186}$ for $\mathbf{10,000}$ Fast / Slow}} \\ \hline
    \end{tabular}
\caption{Timing of different parameter jump types in seconds, from runs on the SIM-MID dataset as described in Table \ref{tab:sim_params}. The total speed up is calculated with a ratio of 10000 Fast samples to any one other jump sample. All results were obtained using single personal computers (PCs) utilizing Intel I9-10900k CPUs with 20 cores at 3.7GHz and 64 GBs of memory at speeds of 3200 MHz. The steps labeled as ``TD" are averages over both the whole step, while all other timings come from the likelihood calls. This is done because the time per likelihood call for TD steps can vary greatly depending on if the model is adding or removing parameters.}  \label{tab:timing_table}
\end{table}
\end{scriptsize}

There are some notable takeaways from this timing data. First, we see \texttt{QuickBurst} yields a $\sim 1.5\times - 2\times$ increase in time per step for jumps that vary shape parameters compared to \texttt{enterprise}. These include $\tau$-Scan, trans-dimensional (TD), and Regular Fisher jumps. Second, we see \texttt{QuickBurst} has an $8\times$ increase in time per step compared to \texttt{enterprise} for Noise jumps, which vary intrinsic pulsar noise. This is due to an increase in the number of inner products required for deterministic signal models. In expanding the likelihood to account for this deterministic model (as shown in eq.~(\ref{eq:logL}), we have a factor of $\frac{N_{F}^{2} + 5N_{F}}{2}$ increase in the number of inner products we need to calculate during these Noise jumps in the M and N matrices seen in both eq.~(\ref{eq: N_matrix}) and eq.~(\ref{eq: M_matrix}), which accounts for the time increase shown.

Despite these increases, we still see a significant speed up thanks to our parameter separation as described in \autoref{ssec:Likelihood}. \texttt{QuickBurst} samples our ``Fast" jumps---which only recalculate $\mathrm{\sigma}_{\mathrm{ki}} (\boldsymbol{\theta}_{p})$---up to $10,000\times$ faster than \texttt{enterprise}. With this advantage over \texttt{enterprise}, we see a factor of $186 \times$ speed up overall for generic GW burst searches in comparison to \texttt{BayesHopperBurst}, even when including RN and CURN timing. If we exclude both CURN and RN, this factor improves up to approximately $204 \times$. 

These times were calculated by averaging over the timing table jump types for \texttt{QuickBurst} and \texttt{BayesHopperBurst}, and computing the average time over 100 samples for each algorithm. Note that the timing for an \texttt{enterprise} likelihood calculation for an equivalent fast jump is on average 100 $\mu s$, which is omitted in the \autoref{tab:timing_table} given that \texttt{BayesHopperBurst} does not have the ability to perform this type of parameter jump.

\section{\label{sec:Analysis_of_Datasets} Analysis of simulated data sets} 

\subsection{Datasets}{\label{ssec:Datasets}}
To test \texttt{QuickBurst}, we simulate four datasets that include a burst signal of varying amplitude. These simulations are based on the Astro4Cast simulations presented in \cite{Astro4Cast} and realistically emulate the 15 year NANOGrav dataset. This means our observing epochs (including gaps) as well as the injected RN and WN in every pulsar are based on measured values. However, to ensure a realistic-sized data set, we do not average the residuals over each observing epoch, and instead retain and analyze all of them. One dataset contains no GW burst signal injection, while the other three datasets include an injected GW burst signal in the form of a parabolic encounter of two SMBHs \cite{Parabolic_enc} each with a mass of $10^9 M_{\odot}$, an impact parameter of $120M_{\odot}$, a sky location of ($\theta_{\rm gw}=\pi/2$, $\phi_{\rm gw}=4.0$) and while varying the source luminosity distance. These four datasets form four signal regimes: a high amplitude (SIM-HIGH), medium amplitude (SIM-MID), low amplitude (SIM-LOW), and no signal case. All datasets contain pulsar noise properties as found in \S5 in \cite{NG15_NoiseBudget}. For these four datasets, we are disregarding the inclusion of a CURN process. See \autoref{sssec:CURN_analysis} for generic GW burst analysis including CURN.

We quantify the signal strength by calculating the SNR of our GW burst injection. The SNR is given by 
\begin{equation}
    \mathrm{SNR} (h) = \sqrt{(h|h)}
\end{equation}
for any signal h. For all three non-zero amplitude datasets, the signal SNR and luminosity distances can be found in \autoref{tab:sim_params}. We calculate the SNR in two cases for each dataset: one case where we only include intrinsic pulsar WN in our covariance matrix, and another case where we include intrinsic pulsar RN and WN in our model. The difference between these SNR values indicates how much the detectability of the burst is affected by RN. The subsequent analysis performed on these four datasets in \autoref{ssec:Analysis} includes both intrinsic pulsar RN and WN. 

\begin{scriptsize}
\begin{table}[ht!]
    \begin{tabular}{|c|c|c|c|}\hline
    \centering
    \texttt{\textbf{Parameters}}& \texttt{\textbf{SIM-LOW}} & \texttt{\textbf{SIM-MID}} 
    & \texttt{\textbf{SIM-HIGH}} \\ \hline \hline 
    luminosity distance  &$ 120 \mathrm{Mpc}\ $ & $60 \mathrm{Mpc}\ $ & $30 \mathrm{Mpc}\ $ \\ \hline 
    WN SNR &$ 4.7\ $ & $9.5\ $ & $19.0\ $ \\ \hline
    WN+RN SNR & $3.9\ $ & $7.8\ $ & $15.7\ $ \\ \hline
    \end{tabular}
\caption{Table consisting of luminosity distance, and signal SNR in the cases of including only intrinsic pulsar WN or both intrinsic pulsar RN and WN. The quoted SNR values were computed with only 24 of 67 NANOGrav pulsars, which correspond to pulsars with more than 10 years of observations. \label{tab:sim_params}}
\end{table}
\end{scriptsize}

\begin{scriptsize}
\begin{table*}[!htb]
    \begin{tabular}{|c|c|c|c|c|c|c|c|c|c|c|}\hline
    \centering
    & \multicolumn{3}{|c|}{Shape parameters}& \multicolumn{7}{|c|}{Projection parameters}\\ \hline \hline 
    \rule{0pt}{7pt} \texttt{\textbf{Parameters}} & $\mathrm{\tau (yrs)}$ & $\mathrm{t_0 (yrs)}$ & $\mathrm{log_{10}(f_0)}$ & $\mathrm{\cos{\theta_{\rm gw}}}$ & $\mathrm{\phi_{\rm gw}}$ & $\mathrm{\psi_{\rm gw}}$ & $\mathrm{A_{+}}$ & $\mathrm{A_{\times}}$ & $\mathrm{\phi_{0, +}}$ & $\mathrm{\phi_{0, \times}}$ \\ \hline
    \rule{0pt}{9pt} \texttt{\textbf{Range}} & $0.2,5.0$ & $0.0,T_{\rm max}$ & $-8.46,-7$ & $-1,1$ & $0,2\pi$ & $0,\pi$ & $10^{-10},10^{-5}$ & $10^{-10},10^{-5}$ & $0,2\pi$ & $0,2\pi$ \\ \hline
    \end{tabular}
\caption{Table of prior ranges used for signal parameters in analysis of the datasets described in \autoref{tab:sim_params}. $T_{\rm max}$ refers to the maximum observation time span in a given dataset with respect to the starting MJD, which is an MJD of $\sim$5855 for these datasets. All angles are in units of radians. \label{tab:param_priors}}
\end{table*}
\end{scriptsize}

Another important note is that the burst epoch in all datasets is at 55000 MJD, which is in the first half of the dataset. This is an arbitrary choice. Many pulsars do not have TOAs around this burst epoch, and will not provide additional significance on GW burst recovery. As such, we truncate the list of pulsars involved in our searches on the datasets described in \autoref{tab:sim_params} to only include pulsars with more than 10 years of data, which is 24 of the 68 pulsars in the NANOGrav 15 year PTA. Note that we only use 67 of these pulsars (see \cite{nanograv_15_gwb, NG15_data} for details). Doing so improves the efficiency of our generic GW burst search, as the model has fewer pulsars to fit a GW burst to. Such a PTA reduction may be possible for real datasets, even without knowing the burst epoch, by investigating the SNR contributions from individual pulsars. See Appendix \ref{sec: ApndxA} for further discussion. However, this does reduce our sky coverage, which is discussed in \autoref{sssec:Nonzero_amp} in further detail.

\subsection{Analysis}\label{ssec:Analysis} 

Now we present the results of the analysis of datasets of varying signal strengths. We first analyze the no signal regime, and then analyze the datasets that have a signal present. Since the parameters of our generic model have no physical meaning, the pipeline's performance is best characterized by how well the waveform of the GW signal is reconstructed. To quantify our signal recovery, we can see how close the injected ($h$) and recovered ($h'$) waveforms are to each other by computing a normalized noise weighted inner product between $h$ and $h'$ over the entire PTA. This match $M$ is defined as

\begin{equation}\label{eq:match_stat}
    \mathrm{M} = \frac{(h|h')}{\mathrm{SNR} (h) \times \mathrm{SNR} (h')}
\end{equation}

where a value of $\mathrm{M} = 1$ is a perfect overlap between $h$ and $h'$, while $\mathrm{M} = 0$ is no overlap between the two waveforms. More details on this statistic can be found in \S2 of \cite{BayesHopperBurst}.

Described in \autoref{tab:param_priors} are the priors assigned to both shape and projection parameters. Given the exploratory nature of our testing, we use wide ranges on many of our priors. We use a uniform prior for all of our signal parameters.

\subsubsection{Zero amplitude dataset}{\label{sssec:zero_amp}}

\begin{figure}[htbp] 
    \centering
    \includegraphics[width = 1\columnwidth]{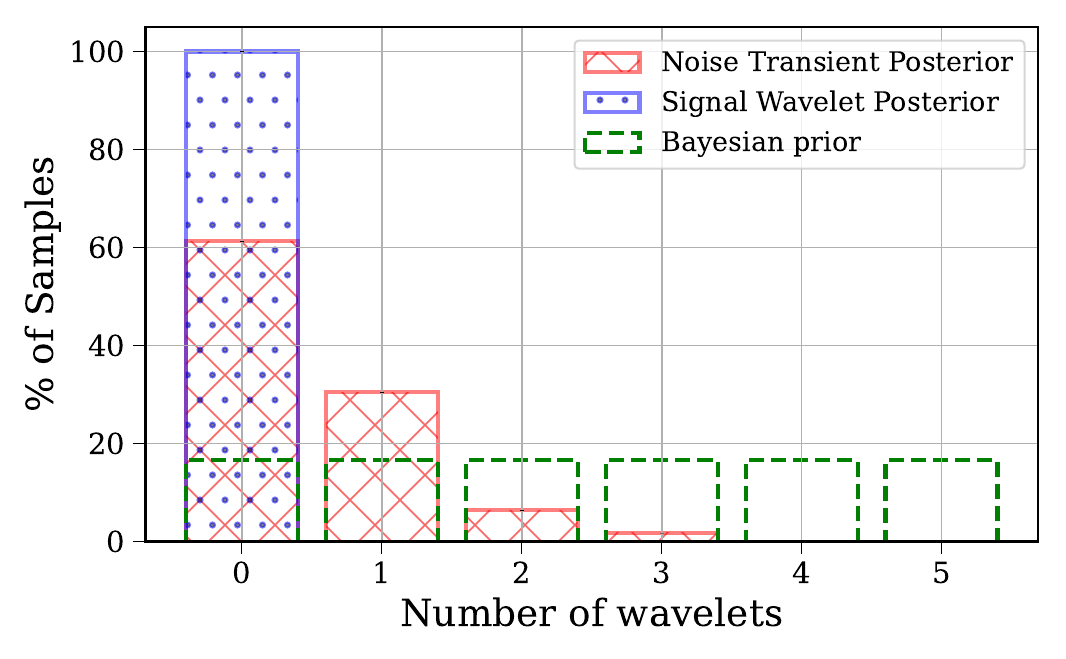}
    \caption{Histogram for recovered noise transients and signal wavelets for our NANOGrav 15-year like dataset with no GW burst injection. We note that the model prefers no signal wavelets to be present in the model at all times in sampling.}
    \label{fig:zero_signal_wavelets}
\end{figure}

To establish a baseline, we have tested \texttt{QuickBurst} on a dataset with no GW burst injection, while keeping all intrinsic pulsar noise as realistic as possible. As expected, our analysis shows preference for no signal wavelets being included in our signal model, in addition to no noise transient wavelets, as shown by the histogram in \autoref{fig:zero_signal_wavelets}. 

\subsubsection{Nonzero signal amplitude datasets}{\label{sssec:Nonzero_amp}}

Here we show our analysis of the three non-zero SNR signal simulated datasets described in \autoref{tab:sim_params}. Given that the waveform shape can vary widely based on the orbit parameters of the two SMBHs, a good indicator for \texttt{QuickBurst}'s ability to recover our signal is to accurately reconstruct the signal waveform after performing searches in our datasets.

\begin{figure*}[htbp!] \label{fig:A4Cast_reconstructions}
   \centering
\subfloat[]{%
\includegraphics[width =1.0\columnwidth]{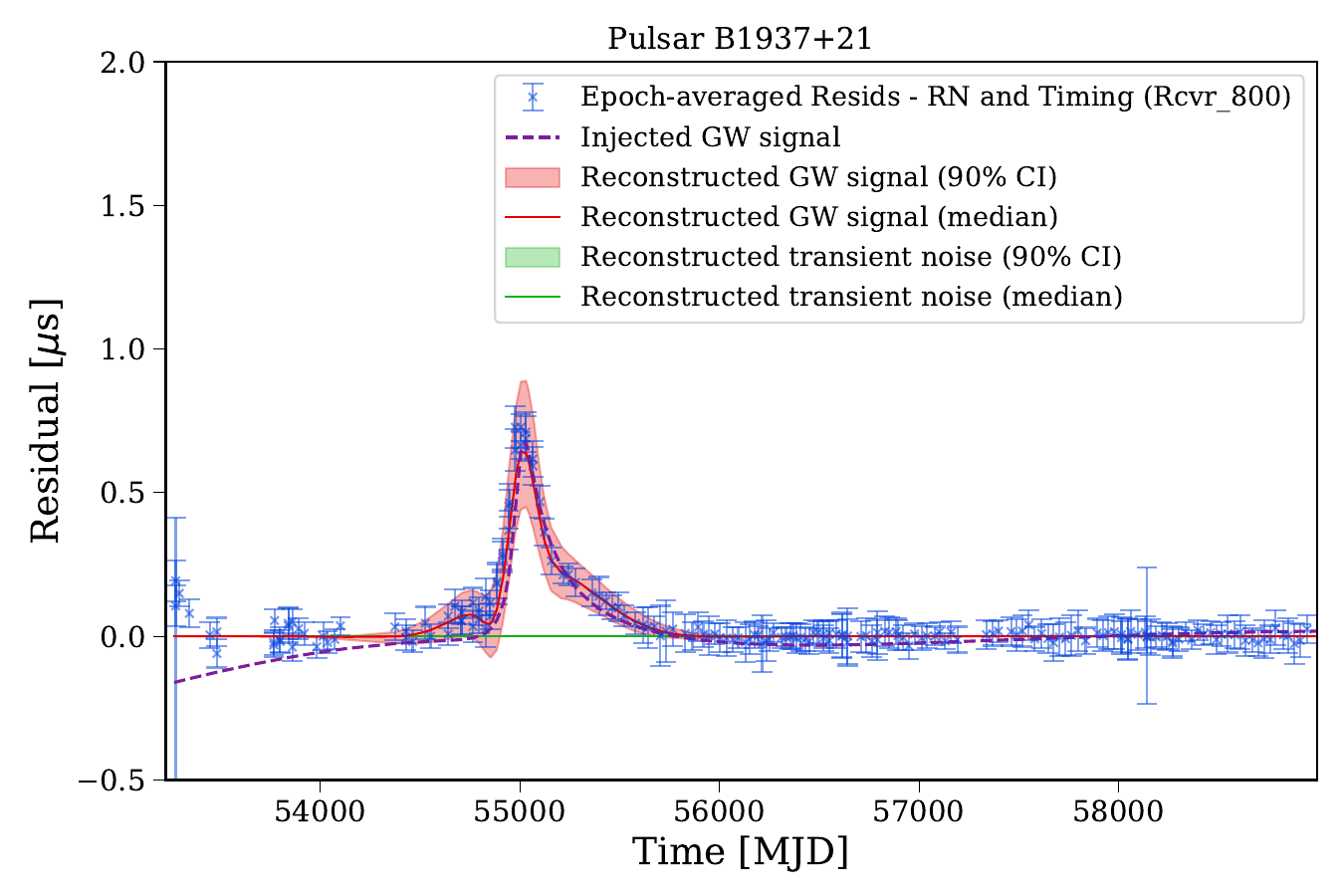}%
\includegraphics[width =1.0\columnwidth]{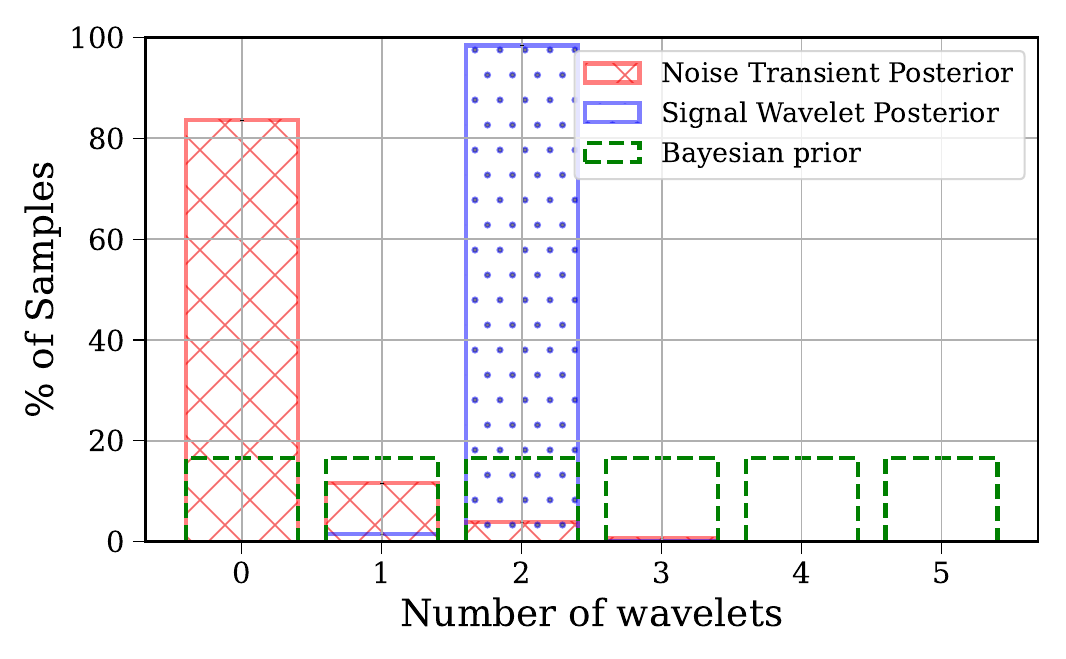}%
}\quad
\subfloat[]{%
\includegraphics[width =1.0\columnwidth]{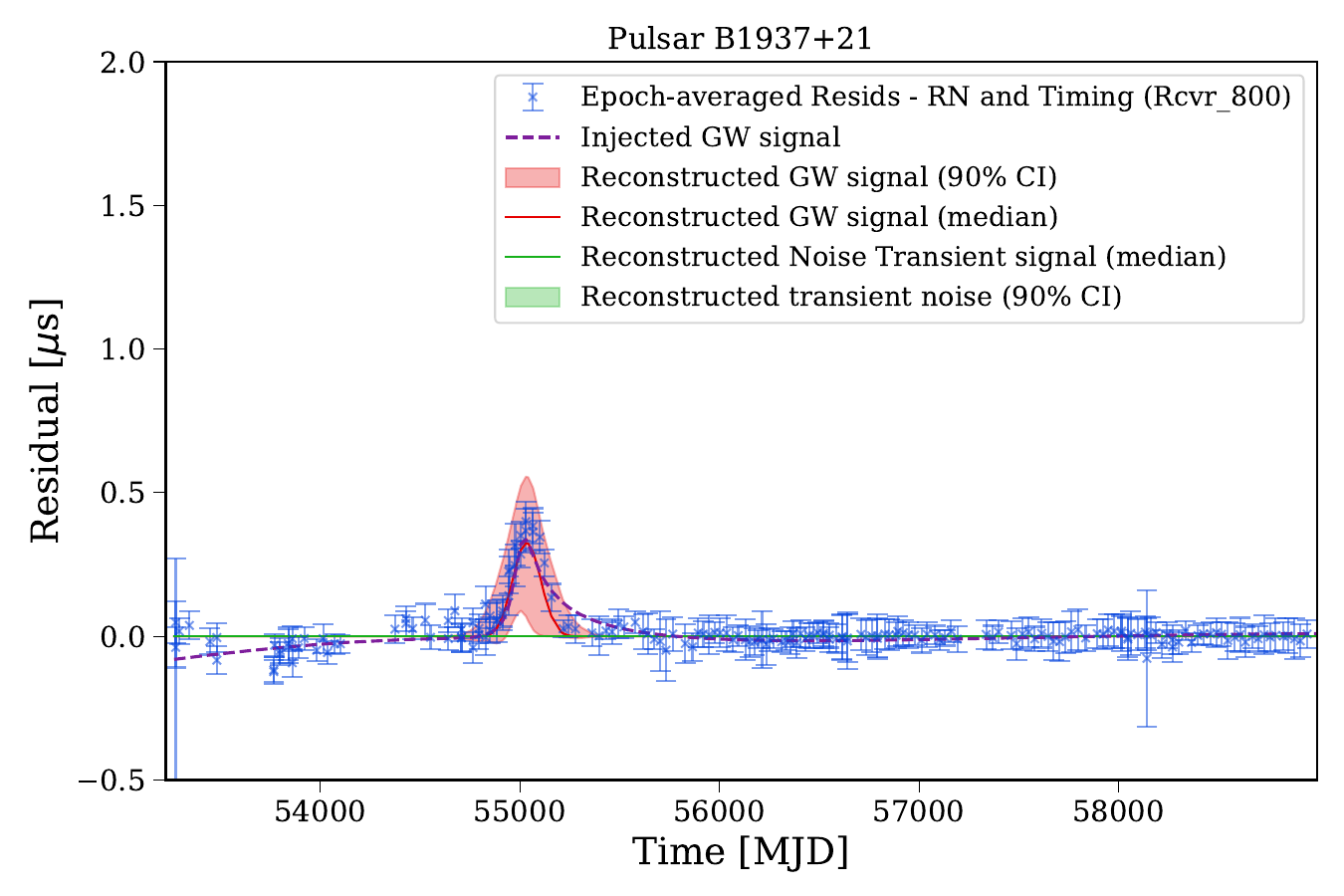}%
\includegraphics[width =1.0\columnwidth]{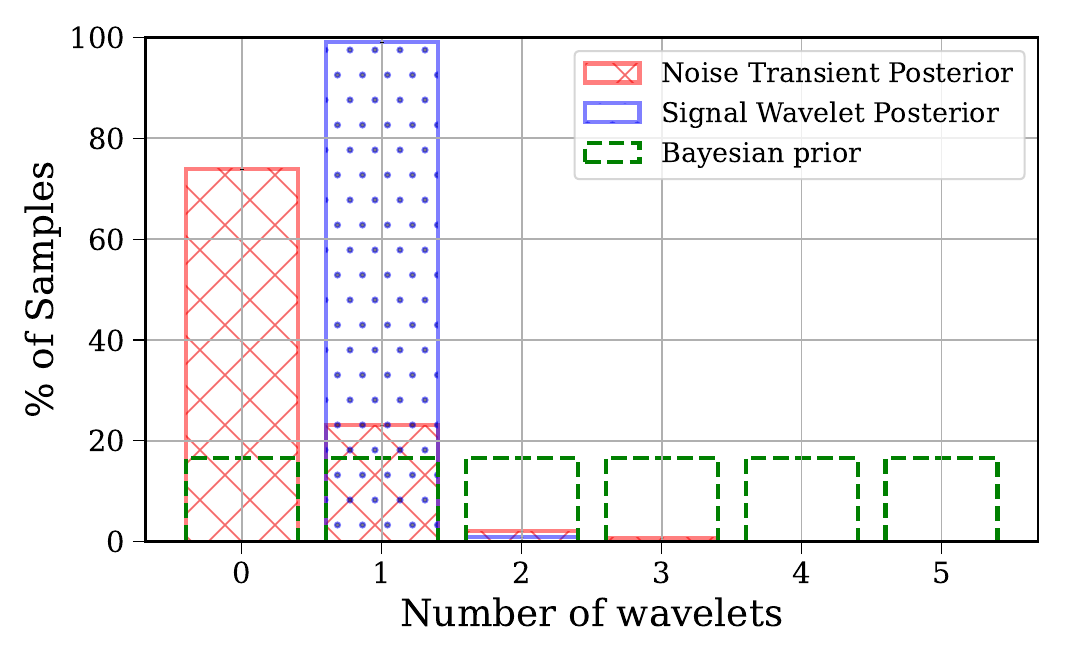}%
}\quad
\subfloat[]{%
\includegraphics[width =1.0\columnwidth]{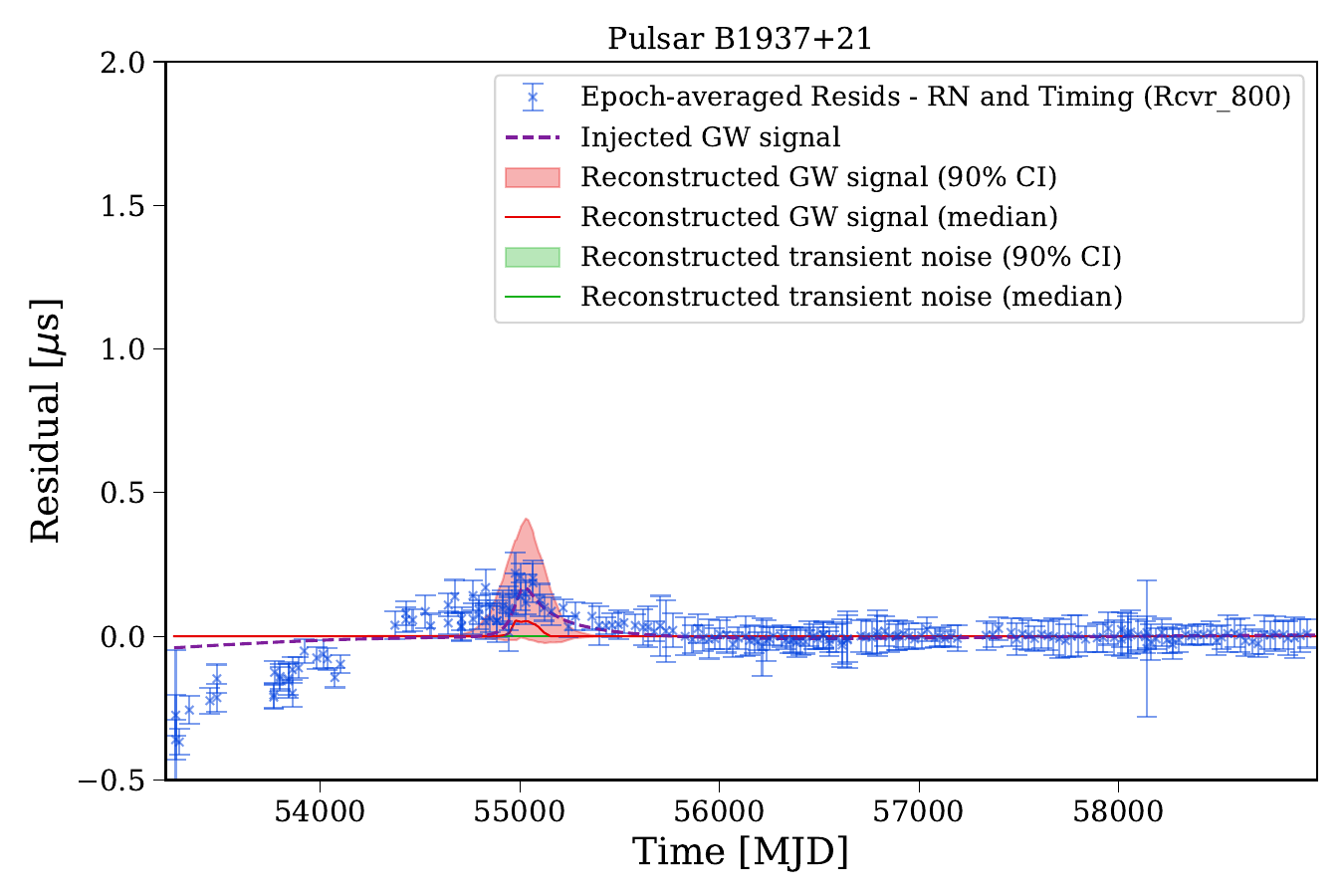}%
\includegraphics[width =1.0\columnwidth]{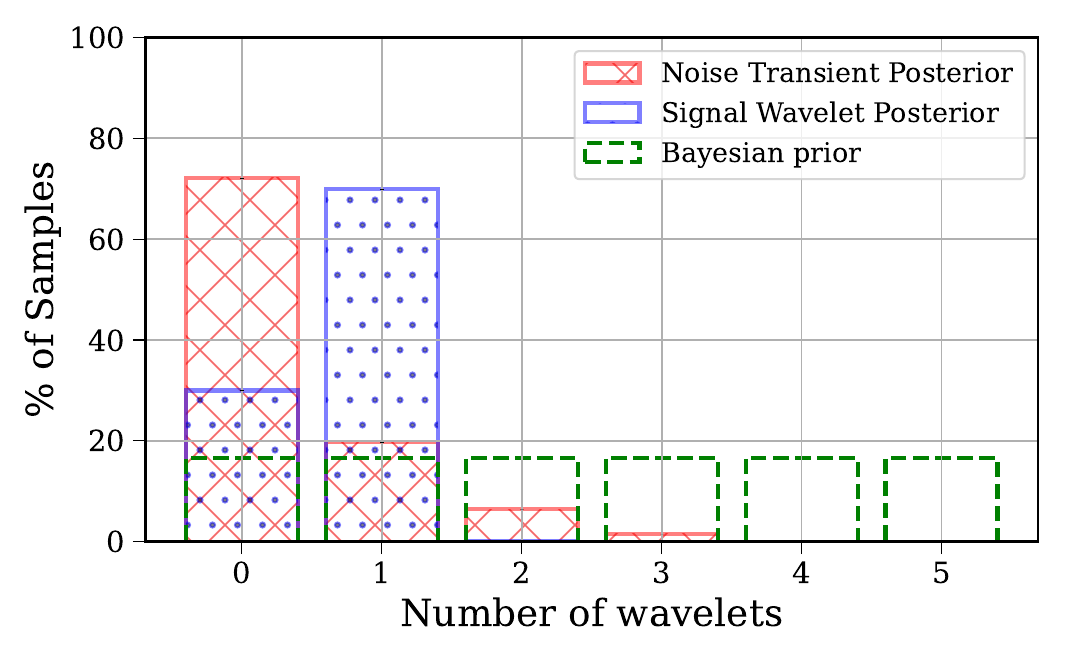}%
}\quad
    \caption{\textbf{(a)} 30 Mpc SMBHB parabolic encounter dataset with an injected signal SNR of 15.7 and $\mathrm{M} = 0.96$. \textbf{(b)} 60 Mpc SMBHB parabolic encounter dataset with an injected signal SNR of 7.8 and $\mathrm{M} = 0.86$. \textbf{(c)} 120 Mpc SMBHB parabolic encounter dataset with an injected signal SNR of 3.9 and $\mathrm{M} = 0.84$. For each dataset, (\textit{left}) Signal reconstruction in B1937+21. (\textit{right}) Number of recovered signal and noise transient wavelets. The waveform reconstructions are created using the La Forge software package \cite{la_forge}.}
    \label{fig:Recovery} 
\end{figure*}

In \autoref{fig:Recovery}, we show the results of our GW signal and noise transient reconstructions in PSR B1937+21. This pulsar has a significant burst amplitude present in all three datasets, with single pulsar GW burst SNR values of 2.8, 1.4, and 0.7 for the SIM-HIGH, SIM-MID, and SIM-LOW datasets respectively. The plots on the left show the epoch averaged residuals for B1937+21 with the RN and best fit timing model parameters ($\boldsymbol{n}_{i} + \boldsymbol{g}_{i}$ and $M_{i}\delta \boldsymbol{\xi}_{i}$ in eq.~(\ref{eq:residuals}), respectively) subtracted out. Additionally, the solid red line is the median waveform reconstruction for the GW burst, and the red shaded region is the 90 \% credible region. The solid green line is the median noise transient reconstruction, and the shaded green region is the 90 \% creible region for the noise transient (if any are present). The plots on the right show the posteriors for the number of signal wavelets and noise transient wavelets for these analyses. The three columns from top to bottom correspond to the three analyzed datasets, SIM-HIGH, SIM-MID, and SIM-LOW, respectively (see Table \ref{tab:sim_params}). 

First, we see that the waveform for the GW bursts in these three datasets are accurately reconstructed. We find that the median reconstructions match the injected waveforms with values $\mathrm{M} = 0.96$, $\mathrm{M} = 0.86$, $\mathrm{M} = 0.84$ for the SIM-HIGH, SIM-MID, and SIM-LOW dataset respectively. We also see the match monotonically increase with GW burst SNR as expected \cite{Bence_waveform}. 

Next, for the recovered signal wavelets, it's been seen in other generic GW burst algorithms that utilize a sine-Gaussian basis (see \cite{BayesWave_SNR}), where the number of recovered wavelets scales with GW burst SNR, as the model can fit for more features in a higher SNR signal. This is consistent with the signal wavelet posteriors in \autoref{fig:Recovery}. Regarding the SIM-LOW dataset, we find that due to the significantly lower signal SNR, convergence took $\sim 3\times$ longer than the SIM-MID and SIM-HIGH analyses. This is a result of marginal differences in the Bayesian likelihood when proposing GW signal wavelets at this SNR. Furthermore, we see there is a significant preference for no noise transient wavelets in all three cases, and that our reconstructions are always consistent with zero. Despite this, there is still a non-negligible percentage of samples at greater than zero noise transients. Having no noise transients in our model is preferred, but having a few noise transients cannot be completely ruled out. This is expected, as white noise fluctuations can be fitted with a small amplitude wavelet. 


\begin{figure}[!htbp]
    \includegraphics[width = 0.75\columnwidth]{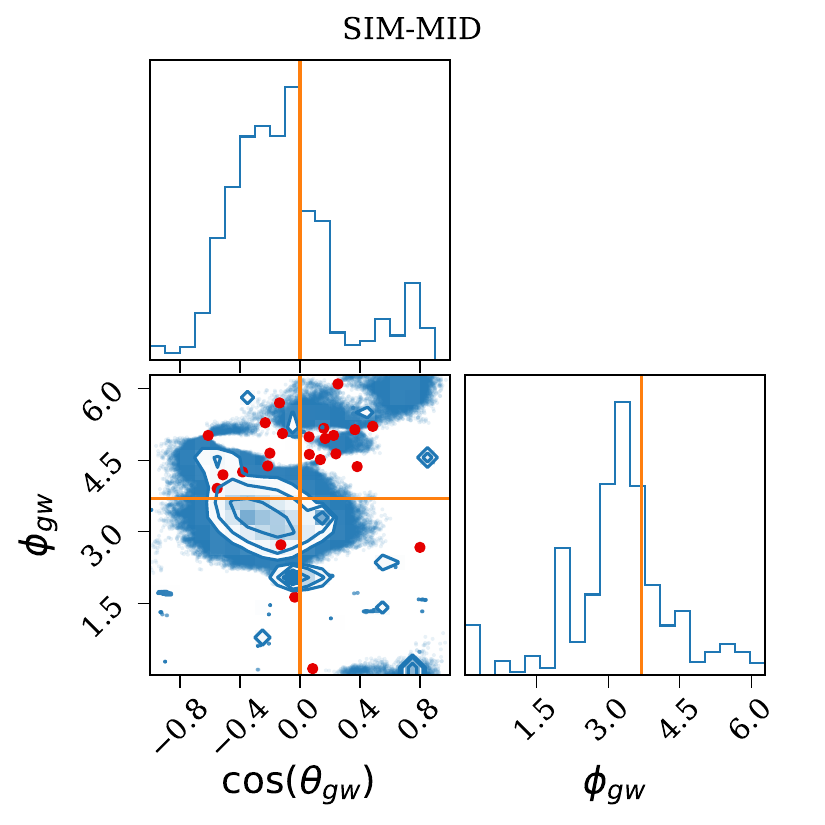}
    \includegraphics[width = 0.75\columnwidth]{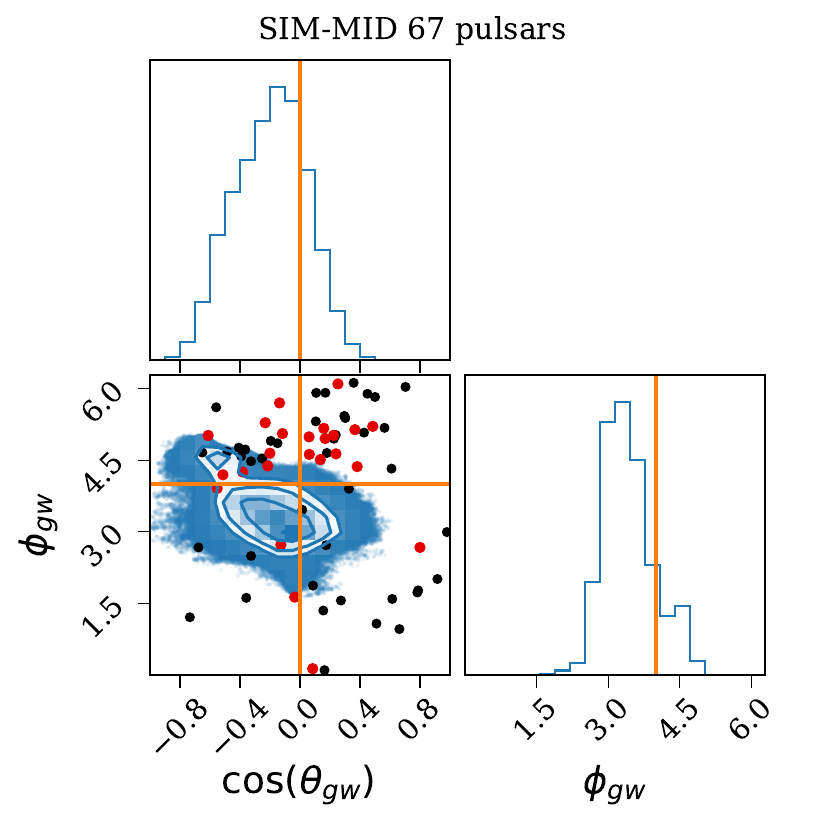}
    \includegraphics[width =0.75\columnwidth]{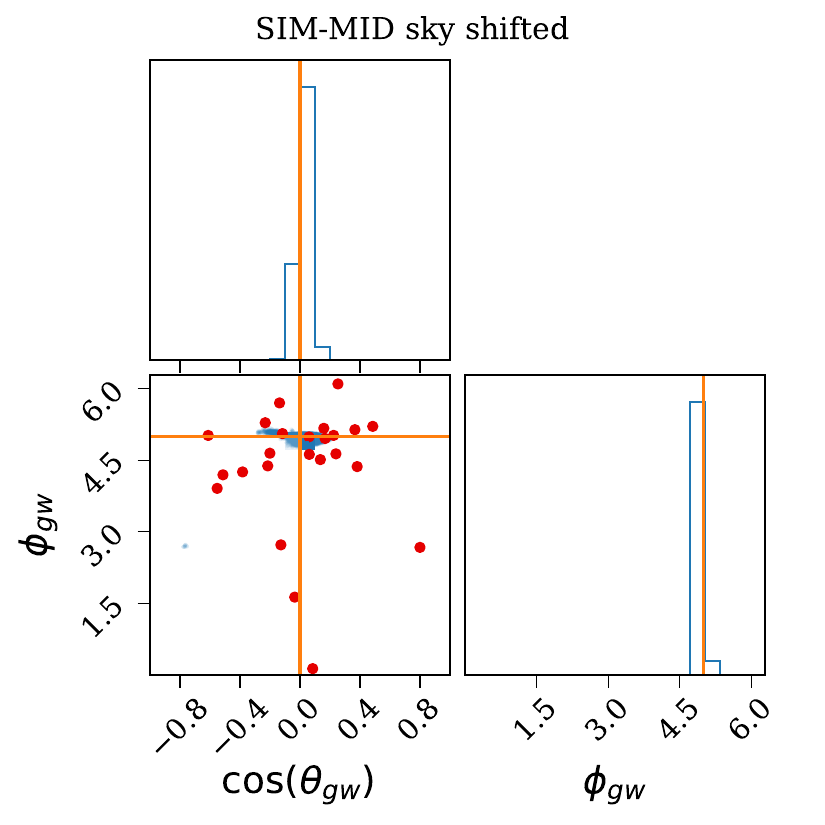}
    \caption{Sky location recovery for a search on the SIM-MID dataset that includes either 24 NANOGrav pulsars, \emph{(top)} and \emph{(bottom)}, or 67 NANOGrav pulsars \emph{(middle)}. The 10 year and higher baseline pulsars are displayed as red dots, while the pulsars with less than a 10 year baseline are displayed as black dots. The lines indicate injected sky location parameters, which are at $\phi_{\rm gw} = 4$ and $\theta_{\rm gw} = \pi/2$ for \emph{(top)} and \emph{(middle)}, but are shifted to $\phi_{\rm gw} = 5.0, \theta_{\rm gw} = \pi/2$, and $\psi_{\rm gw} = -0.3$ for \emph{(bottom)}. All angles are in units of radians.}
    \label{fig:SkyCorner}
\end{figure}

We also test that \texttt{QuickBurst} can properly recover the sky location of a GW burst. In \autoref{fig:SkyCorner} we show the sky recovery results of three different runs. In the top two plots, we plot the sky location recovery for a search on the SIM-MID dataset, both where we include only our longest timed pulsars (top figure), and where we include all the pulsars in the NANOGrav 15 year PTA (middle figure). We then display the sky location recovery for a run on the SIM-MID dataset, but with a shifted GW burst sky location (bottom figure, referred to as the "SIM-MID sky shifted" dataset). 

For the original SIM-MID search, the burst is only constrained to around a quadrant of the sky. We see, however, that including all NANOGrav 15 year pulsars (as shown in the middle figure) doesn't provide additional constraints on our sky location recovery for this particular GW burst signal. The pulsars labeled by black dots are the additional pulsars we are adding, but there's limited data for those pulsars around the burst epoch, despite lying in a more sensitive sky region. It's also important to note that GW burst signals of this type have a finite width. While there are small changes in the posterior between the top two panels, the overall shape is the same between the 24 pulsar case and the 67 pulsar case. This is expected behavior given that these additional pulsars contribute little new GW burst information.

To explore whether this is a feature of the SIM-MID dataset or our pipeline, we can simply change our sky location of the burst to have overlap with more of our longer timed pulsars. Moreover, we ensure that the total GW burst SNR is close to the signal SNR in the SIM-MID dataset. In doing so, we create a new dataset that is almost identical to the SIM-MID dataset, except the burst location is now at $\phi_{\mathrm{gw}} = 5.0  \ \mathrm{rad}, \theta_{\mathrm{gw}} = \pi/2$, and $\psi_{\rm gw} = -0.3  \ \mathrm{rad}$. Based on the bottom figure in \autoref{fig:SkyCorner}, we clearly recover the injected burst location with a shifted burst location. In comparison with the sky location recovery efforts (as shown in the top and middle panels of \autoref{fig:SkyCorner}), we see that having better sky coverage allows us to better recover both $\theta_{\mathrm{gw}}$ and $\phi_{\mathrm{gw}}$, so these sky recovery differences are features of the datasets, and not a limitation of the sampler. We will discuss the issue of sky coverage and GW burst sensitivity more in Appendix \autoref{sec: ApndxA}.

\subsubsection{Common red noise effects}\label{sssec:CURN_analysis}
The analysis we have discussed so far does not include a common uncorrelated red noise, or CURN. This shows the effectiveness of \texttt{QuickBurst} in a simpler context, but as we have seen from the NANOGrav 15 year GWB results \cite{nanograv_15_gwb}, there is strong evidence for a correlated GWB. While it is beneficial to analyze the impacts of a correlated RN process on GW burst analysis, it is not expected that adding in Hellings-Downs (HD) correlations produces significant additional covariance with our wavelet model (see \autoref{sec: ApndxB} for more details). Therefore, we will show \texttt{QuickBurst} can disentangle a GW burst from a CURN process.

To isolate the effects a CURN process would have on our dataset, we test on a simple simulated data set of 20 pulsars with evenly sampled data that includes simulated pulsar WN, CURN, and a GW burst signal from a parabolic SMBH encounter at a sky location of $\phi_{\mathrm{gw}} = 3.7  \ \mathrm{rad}, \theta_{\mathrm{gw}} = \pi/2$, and $\psi_{\rm gw} = 0.2  \ \mathrm{rad}$ (unlike the results in \autoref{fig:Recovery} which used a more realistic dataset made to resemble the NANOGrav 15yr dataset). This dataset consists of a CURN injection with an amplitude of $A=6.0 \times 10^{-15}$, a spectral index of $\gamma = 13/3$, and $1 \mu s$ of intrinsic pulsar WN (which sets the CURN process to be larger than the WN at the lowest frequencies).

\begin{figure}[!htp]
    \subfloat[]{%
    \includegraphics[width = 1.0\columnwidth]{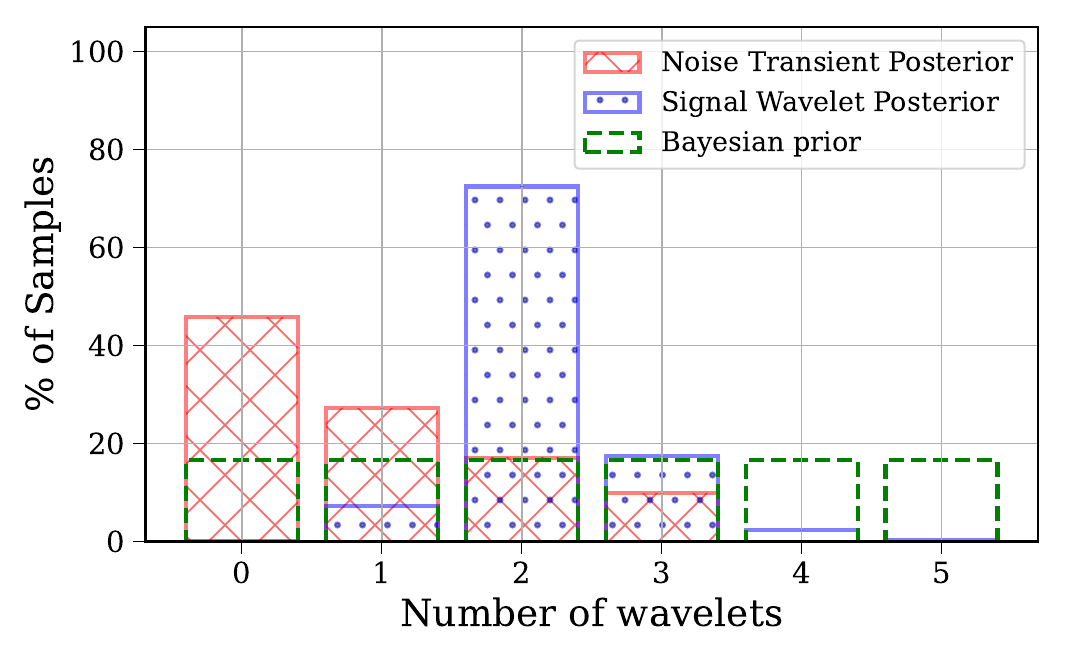}
    }\quad
    \subfloat[]{%
    \includegraphics[width = 0.90\columnwidth]{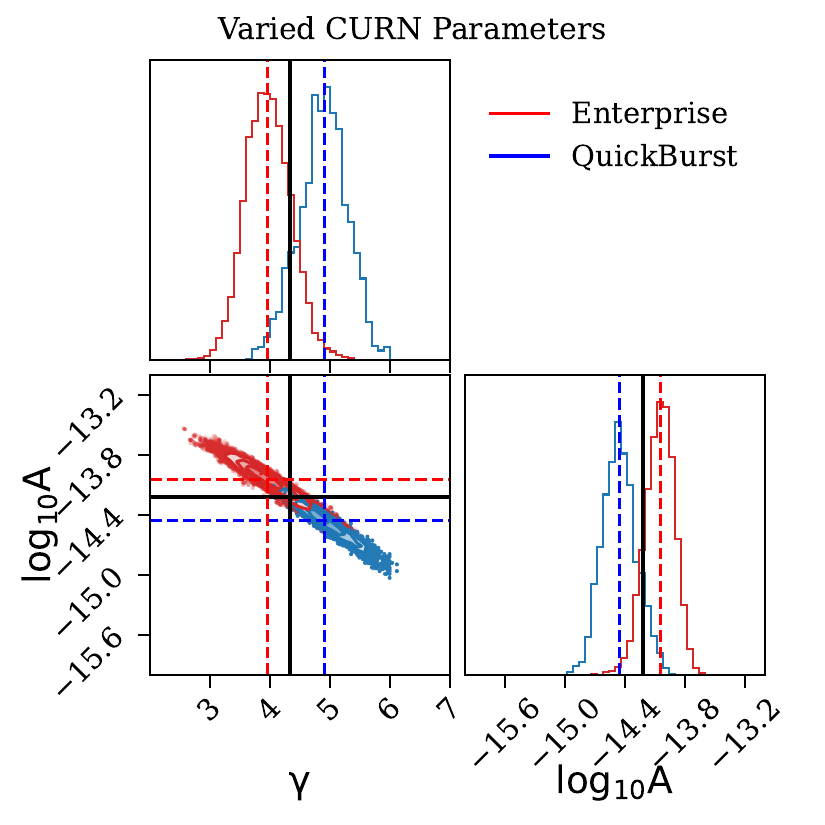}
    }\quad
    \caption{Plots using a dataset with 20 pulsars and a CURN of amplitude $A=6.0 \times 10^{-15}$. CURN parameters were allowed to vary. \textbf{(a)} Histogram of recovered signal and noise transient wavelets as found in \texttt{QuickBurst} while simultaneously searching over CURN spectral parameters.  \textbf{(b)} Corner plot of recovered CURN power law parameters from both the \texttt{QuickBurst} run that generated the top plot (blue), and a search using \texttt{enterprise} for only CURN and other noise intrinsic pulsar (red). The injected values were $\gamma=4.33$ and $\mathrm{log_{10}A = -14.22}$ overlay-ed as black lines, while the dotted lines correspond to the median recovered values for each run. \texttt{QuickBurst} recovered median values of  $\mathrm{\gamma=4.90}$ and $\mathrm{log_{10}A = -14.45}$, while the search with \texttt{enterprise} yielded  median values of $\gamma=3.96$ and $\mathrm{log_{10}A = -14.05}$. Both of these recoveries are consistent with the injected signal within a $65 \%$ credible region.}
    \label{fig:CURN_wavelet_hist}
\end{figure}

First, we perform a search with \texttt{QuickBurst} where we fix our CURN process to the injected CURN values. In doing so, we recover a significant portion of our injected signal, boasting a recovered SNR of $6.4$ (injected SNR of 6.8) and a match of $\mathrm{M = 0.82}$. We then perform a run on the same dataset where we simultaneously vary both wavelet parameters and CURN spectral parameters, which is displayed in \autoref{fig:CURN_wavelet_hist} (a). Results indicate a match of $\mathrm{M = 0.79}$ and a recovered SNR of 10.2 (injected SNR of 9.0). Furthermore, in comparison to histograms shown in \autoref{fig:Recovery}, the distribution of signal wavelets and noise transient wavelets are wider. This is likely due to covariance between the CURN process at overlapping frequencies, which can result in some of the GW burst power being absorbed into the CURN model. This suggests that even in strongly red noise dominated data, it is possible to disentangle the signal. There may be additional complications on real data in the future, where additional methods may help to better disentangle these two signal types, if a GW burst is present. 

It is also informative for future searches to see the effects of excluding a GW burst from our overall signal model. To analyze these effects, we perform a search using \texttt{enterprise} on our simple dataset where only CURN is being modeled (and varied), and compare to a search using \texttt{QuickBurst} where both CURN and our GW signal wavelet parameters are varied. \autoref{fig:CURN_wavelet_hist} (b) shows the posterior distribution of the CURN amplitude (A) and the sepctral index ($\gamma$) from these two runs. We can see that the difference in posteriors illuminates a bias in the recovery of a CURN process.

In the \texttt{QuickBurst} search, given that there is overlapping frequency content between a CURN signal and a GW burst, the GW burst signal model can absorb some of the CURN power at the overlapping frequencies. This bias sharpens the CURN spectrum, raising the spectral index and lowering its amplitude at higher frequencies. Inversely, in the \texttt{enterprise} search, our recovered CURN parameters are biased towards a lower spectral index and a higher amplitude. This is a result of the GW burst having higher frequency content, but still having a larger SNR than the WN present. This high frequency GW burst content flattens the CURN spectrum while increasing its overall amplitude. This effect means that we will have to balance using fixed vs varied runs on real datasets to try and most accurately recover any signals that may be present.

\section{\label{sec:Conclusion} Conclusions} 

We have developed a pipeline that can search for generic GW bursts in PTA datasets using a basis of Morlet-Gabor wavelets. This method has been demonstrated to improve search efficiency by $\sim 200 \times$. \texttt{QuickBurst} is capable of recovering detectable signals in a variety of signal strength regimes, while buried in a realistic simulated PTA with intrinsic pulsar RN, intrinsic pulsar WN, and non-uniform sampling/sky coverage. \texttt{QuickBurst} can also be used to efficiently find noise transients in PTA data. Mitigating such noise transients can have a positive effect on other GW analyses, as it can help make sure the presence of such transients do not bias the GW parameter recovery.

With this pipeline developed, it is now computationally feasible to search agnostically for any kind of gravitational wave burst signal in the newest PTA datasets, like the NANOGrav 15 yr dataset, the EPTA DR2 dataset, or the PPTA DR3 dataset \cite{NG15_data, EPTA_DR2_data, PPTA_DR3_data}. \texttt{QuickBurst} will also allow us to scale this analysis to future combined datasets, like the upcoming IPTA DR3 \cite{3P+_comparison}. Future work includes implementing the ability to simultaneously searching for a generic GW burst signal alongside a common correlated RN process, which would take into account the most recent results found in \cite{nanograv_15_gwb, epta_dr2_gwb, ppta_dr3_gwb, cpta_dr1_gwb}. Future work also includes expanding this approach to include radio-frequency-dependent wavelets, which can be useful for modeling transient changes in the dispersion measure or other chromatic effect (for a review, see \cite{NG15_NoiseBudget}). Lastly, DMX is a flexible model that can easily fit features that are not in fact DM variations. This means that DMX can make the covariance between the DM and burst models worse than it intrinsically is, especially on short time scales. Future work may include investigating this covariance through modeling DM as a Gaussian process rather than epoch-by-epoch offsets \cite{Larsen_DMGP, IPTA_DMGP}.

\section{\label{sec:Acknowledgements} Acknowledgements} 
We would like to thank Xavier Siemens and Jerry Sun for their insightful discussions throughout the project. This NANOGrav project receives support from National
Science Foundation (NSF) Physics Frontiers Center award
Nos. 1430284 and 2020265. This work was also partly
supported by the George and Hannah Bolinger Memorial Fund
in the College of Science at Oregon State University. 

\appendix

\section{\label{sec: ApndxA} Per pulsar generic GW burst SNR predictions}

\begin{figure*}[htbp!]
    \centering
    \includegraphics[width = 1.0\columnwidth]{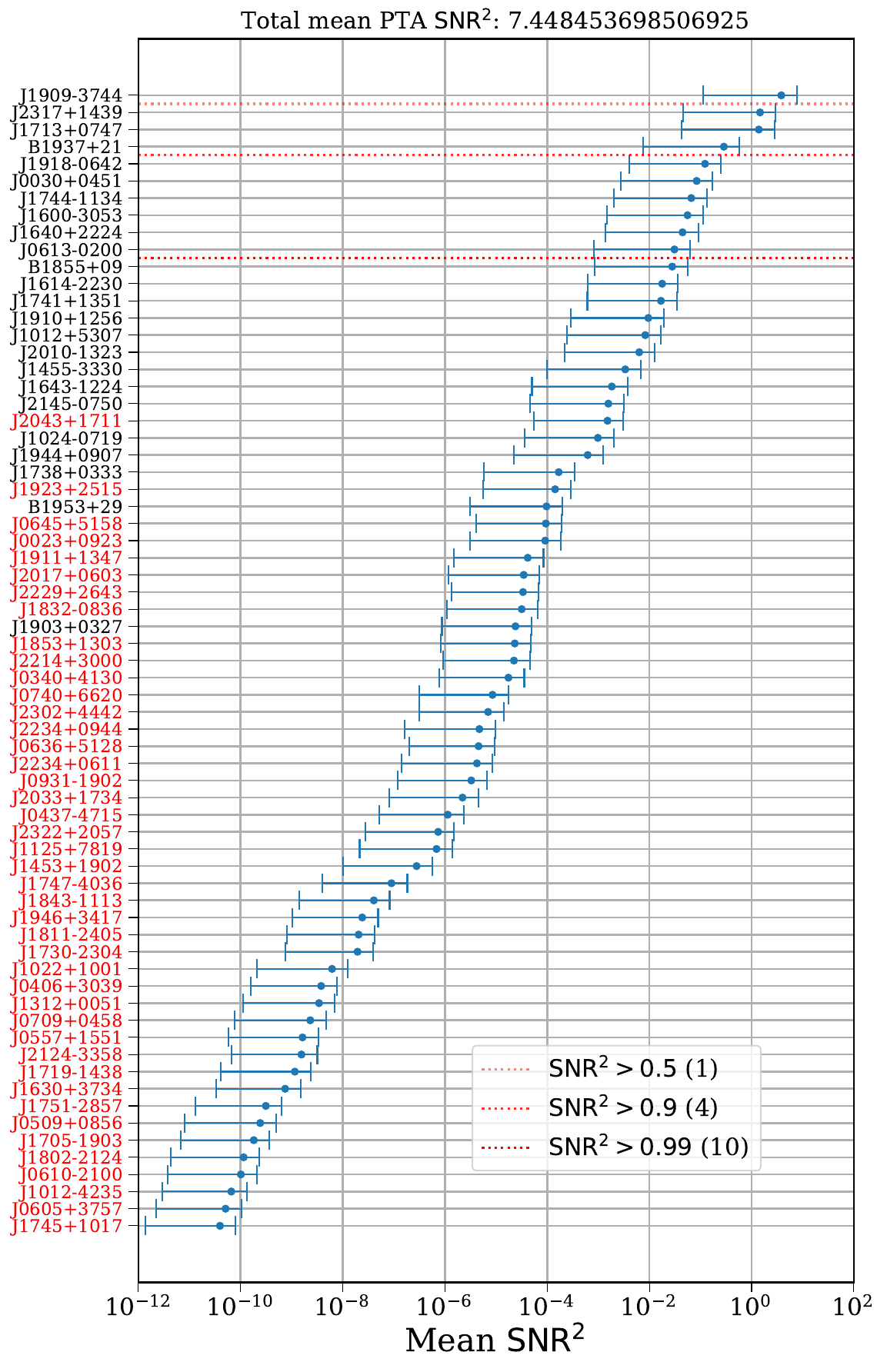}
    \includegraphics[width = 1.0\columnwidth]{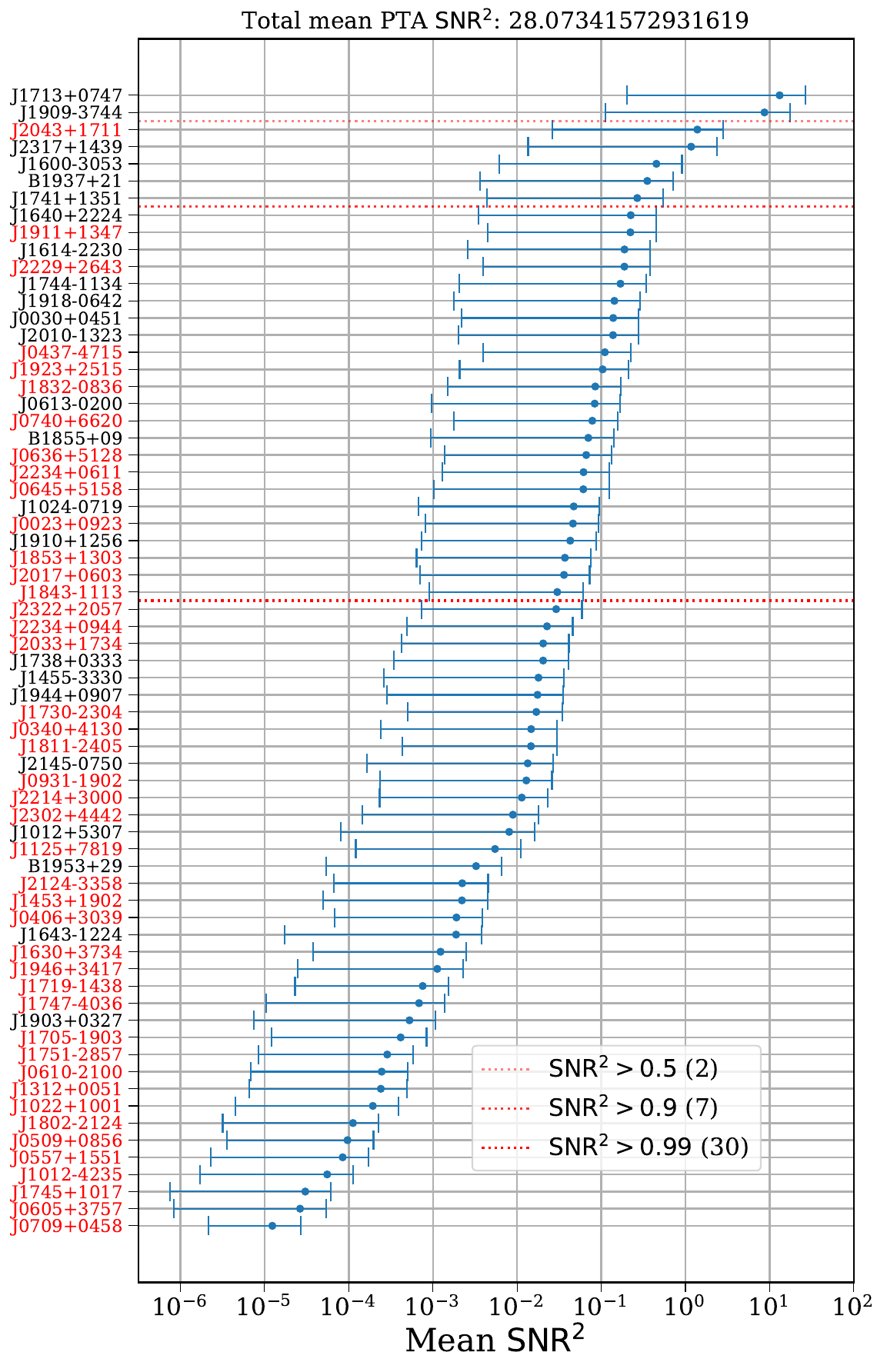}
    \caption{Sky averaged burst $\mathrm{SNR}^{2}$ of pulsars in the SIM-LOW dataset. The mean SNR for the full PTA is listed at the top of each plot for each set of 10000 realizations. (\textbf{\emph{left}}) Mean $\mathrm{SNR}^{2}$ for all pulsars over 10000 realizations of randomized burst locations in the sky at a fixed burst epoch of $\mathrm{MJD} = 55000$. (\textbf{\emph{right}}) Mean $\mathrm{SNR}^{2}$ for all pulsars over 10000 realizations for both randomized burst locations on the sky and randomized burst epochs. Pulsars in red indicate the pulsars with less than 10 years of data, which are pulsars that were dropped from our analyzed datasets described in \autoref{sec:Analysis_of_Datasets}. The dashed red lines indicate the cumulative $\mathrm{SNR}^{2}$ at $50 \%$, $90 \%$ and, $99 \%$, and in parenthesis are the number of pulsars that are included in the cumulative sum.}
    \label{fig:SNR_skyaverage_simlow}
\end{figure*}

In the search for a stochastic background, the SNR of the HD correlations scales with the number of pairs in our PTA \cite{Xavi_scalinglaws}, which motivates using all the available pulsars in the analysis. However, for a generic GW burst, we need a large enough observing time span for any given pulsar to resolve burst features, and pulsars whose noise properties are understood well enough to be subtracted out as accurately as possible. In addition, removing pulsars decreases the computational cost of these searches, giving us more time to resolve the features in the most significant pulsars. In summary, it may be more efficient to remove pulsars with an observation time span less than the maximum width (given by $\tau$) of a GW burst allowed in a GW burst search. These short timed pulsars will therefore be uninformative regarding the morphology of a GW burst, and can bias $\mathrm{SNR}$ values. These pulsars will then contribute minimally to the total GW burst $\mathrm{SNR}$ \cite{Speri_etal_droppingPulsars}. For this reason, the maximum width on the GW signal wavelets is set to $\tau = 5$ years for analysis displayed in \autoref{sssec:Nonzero_amp}.



To test for uninformative pulsar SNR contributions, we can calculate the $\mathrm{SNR}^{2}$ for each pulsar in the SIM-LOW dataset, but not limited to pulsars with the longest observation time-span. We compute these values over 10,000 realizations and average them in two different cases: (1) the burst epoch is fixed to MJD 55,000 (to match the analysis described in \autoref{sssec:Nonzero_amp}), while the GW burst sky location is randomized between each realization, and (2) where both the GW burst epoch and sky location are randomized between each realization. Assuming there is a GW burst present, the second case is the most agnostic, as in real datasets, we will not know when or where a GW burst will present itself. We then take the mean $\mathrm{SNR}^{2}$ value over all 10,000 realizations in each pulsar along with the $65 \%$ credible regions for both cases, and these are plotted in \autoref{fig:SNR_skyaverage_simlow}.

In the fixed epoch case (left figure), the pulsars which have an observing time less than 10 years of data (highlighted in red) show a clear trend to lower $\mathrm{SNR}^{2}$ values. This directly results from the GW burst epoch occurring at a time when those pulsars have no observational data and miss most GW burst features. In the random epoch case (right figure), the trend is less clear. Generally, there is a higher density of longer observed pulsars having larger $\mathrm{SNR}$ contributions. However, there are several exceptions to this trend (namely pulsar $\mathrm{J2043+1711}$) that do not allow us to simply cut pulsars from the analysis based on observation time span. In this case, there are 30 pulsars whom contribute to $99 \%$ of the total GW burst SNR, but not all of these 30 pulsars are ones with a greater than 10 year baseline. Therefore, simply dropping pulsars from the analysis is not an advisable solution for additional computational efficiency. 

However, the expected SNR calculation presented here provides a compelling alternative, which can be used to at least halve the number of pulsars to analyze without losing any significant sensitivity. Such an approach will make future analyses of real data even more efficient. This may also suggest performing future searches with a tiered approach. One could start by analyzing a dataset with the best $\sim 10$ pulsars (which would encapsulate $> 90 \%$ of the SNR), and then perform follow up searches where more pulsars are added. This tiered approach will increase search efficiency by reducing the data volume being analyzed, but at the cost of GW burst sky location sensitivity, as shown in \autoref{fig:SkyCorner}.

\section{\label{sec: ApndxB} HD correlations covariances with generic wavelet modeling}

Recently, the first evidence for the stochastic GWB has been found in PTA data \cite{nanograv_15_gwb, epta_dr2_gwb, ppta_dr3_gwb, cpta_dr1_gwb}. This motivates considering how HD correlations in the common red noise could impact GW burst analysis. In this analysis, we first consider the spectral characterizations of these common signals to see how significant we expect the impact to be.

As seen in the results from various PTAs, the stochastic GWB has it's highest contributions in the lower frequency bins based on free spectrum analysis, which are frequencies ranging from $\sim 1-10$ nHz. Because these GW burst searches are intended to search for signals that persist for less than the time span of a dataset, it is not expected that the spectral power from the GWB will significantly overlap with the peak spectral power of a GW burst. An example can be seen in \autoref{fig:Recovery}, as the wavelet model prefers a frequency between $f_{0} \sim 7-9$ nHz, which does not overlap with the highest power contributions of a stochastic GWB with the most recent spectral characterization of the GWB.

\begin{figure}[!htp]
    \subfloat[]{\includegraphics[width = 1.0\columnwidth]{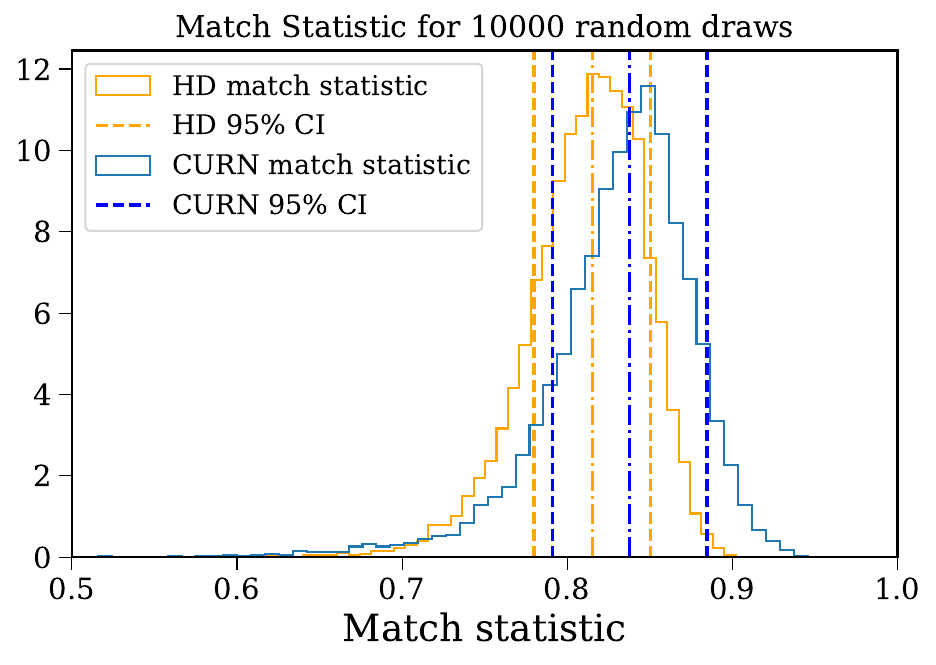}}
    \quad
    \subfloat[]{\includegraphics[width = 1.0\columnwidth]{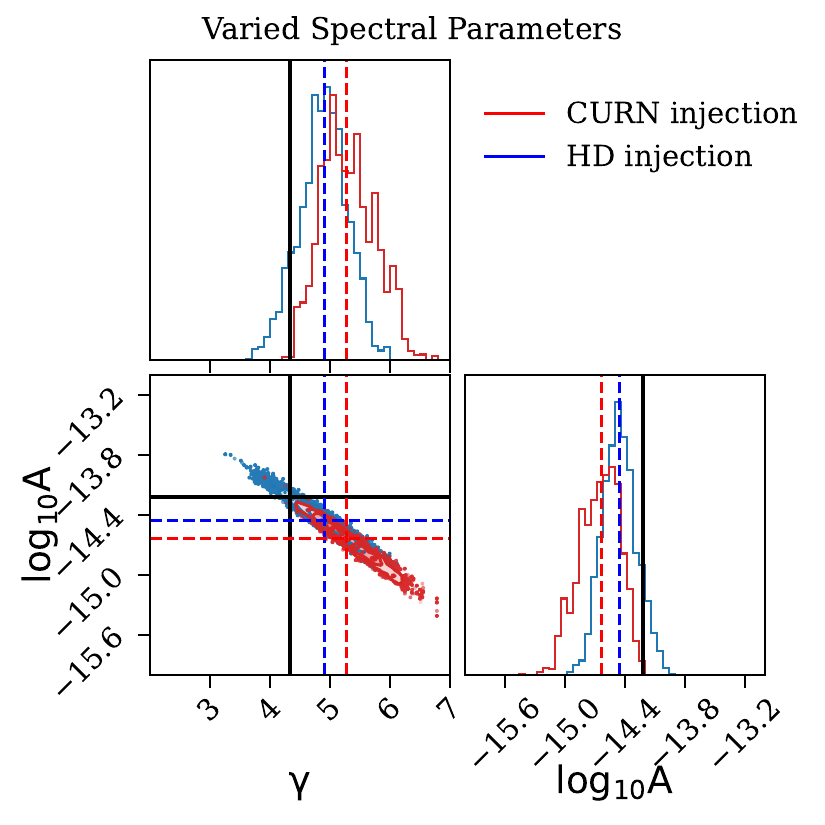}}
    \quad
    \caption{Comparison between two datasets consisting of identical GW burst signals and intrinsic pulsar noise, but only differing in the common noise. One dataset contains an HD correlated signal and the other contains a CURN signal, yet both have the same spectral parameters of $A = 6 \times 10^{-15}$ and $\gamma = 13/3$. \emph{(a)} Distribution of the match statistic from identical analyses on both datasets. \emph{(b)} Corner plot of spectral parameter recovery between both datasets. The black crosshair indicates the injected parameter values for both datasets, while the dashed lines represent the median recovered vales for each dataset.}
    \label{fig:CURN_HD_comparison}
\end{figure}

To test this, we create a simulated dataset that is identical in construction to the dataset used to analyze the inclusion of CURN (as shown in \autoref{fig:CURN_wavelet_hist}), with the only difference being including a correlated RN process with HD correlations versus CURN, both with the same spectral parameters. We then perform identical searches with \texttt{QuickBurst} on both datasets. In both cases, we allow the common process in both analyses vary, full knowing \texttt{QuickBurst} only has the ability to model both common processes as CURN.

In \autoref{fig:CURN_HD_comparison}, we see the results of this analysis. In the top panel, we compute the match statistic for 10,000 randomly drawn samples from the chain returned by \texttt{QuickBurst} for each dataset, and plot the distribution for each. For the dataset containing CURN, the median value of the match statistic is $M = 0.84 \pm 0.05$, and for the median for the dataset containing HD is $M = 0.82 \pm 0.04$. We see that the match statistic distributions are consistent within $1 \sigma$ of each other. The slight decrease in the match statistic distribution for the dataset with an HD GWB is insignificant due to significant variations between noise realizations in simulations. Therefore, this shows a negligible increase in covariance between the inclusion of HD correlations and our wavelet model.

Moreover, in the bottom panel of \autoref{fig:CURN_HD_comparison}, we see the spectral recovery for the common signals of each dataset. The black crosshair signifies the injected signal parameters for each, and the dashed lines represent the median recovered spectral parameters for each run. For the CURN dataset, the median recovery for the spectral parameters are $A = 3.5 \times 10^{-15}$ and $\gamma = 4.9$. For the HD dataset, the median recovery for the spectral parameters are $A = 2.5 \times 10^{-15}$ and $\gamma = 5.3$. Given that there are minimal changes in the posteriors for the spectral parameters, this implies that the model for HD correlations absorbs little additional power than the CURN model has shown to absorb from the GW burst that is present. While we can't conclusively say for sure that a GW burst of any morphology won't have significant covariance with an HD correlated GWB based on this one example, it is promising that adding in these correlations doesn't produce any issues with our wavelet model. Similar simulations can easily validate this for other waveforms with different time-frequency content in the future.

\bibliography{main}{}
\bibliographystyle{unsrt_et_al_15-3author_notitle}
\end{document}